\documentclass[12pt,preprint]{aastex}

\newcommand{\simlt}{\lower.5ex\hbox{$\; \buildrel < \over \sim \;$}}

\slugcomment{draft date: \today}

\begin{document}

\title{Absolute Calibration of the Infrared Array Camera on the
Spitzer Space Telescope}

\author{William T. Reach\altaffilmark{1}}

\author{S. T. Megeath\altaffilmark{2}}

\author{Martin Cohen\altaffilmark{3}}

\author{J. Hora\altaffilmark{2}}

\author{Sean Carey\altaffilmark{1}}

\author{Jason Surace\altaffilmark{1}}

\author{S. P. Willner\altaffilmark{2}}

\author{P. Barmby\altaffilmark{2}}

\author{Gillian Wilson\altaffilmark{1}}

\author{William Glaccum\altaffilmark{1}}

\author{Patrick Lowrance\altaffilmark{1}}

\author{Massimo Marengo\altaffilmark{2}}

\author{Giovanni G. Fazio\altaffilmark{2}}

\altaffiltext{1}{{\it Spitzer} Science Center, MS 220-6, 
California Institute of Technology,
Pasadena, CA 91125}
\altaffiltext{2}{Harvard-Smithsonian Center for Astrophysics, 60 Garden Street, 
Cambridge, MA 02138}
\altaffiltext{3}{Radio Astronomy Laboratory, 601 Campbell Hall, University of California
at Berkeley, Berkeley, CA 94720}

\email{reach@ipac.caltech.edu}

\begin{abstract}
The Infrared Array Camera (IRAC) on the Spitzer Space Telescope is
absolutely calibrated by comparing photometry on a set of A stars
near the north ecliptic pole to predictions based on ground-based 
observations and a stellar atmosphere model. The brightness of 
point sources is calibrated to an accuracy of 3\%, relative to
models for A star stellar atmospheres, for observations 
performed and analyzed in the same manner as the calibration stars.
This includes corrections for location of the star in the array and
the location if the centroid within the peak pixel. 
Long-term stability of the IRAC photometry was measured by monitoring 
the brightness
of A dwarfs and K giants (near the north ecliptic pole) observed several
times per month; the photometry is stable to 1.5\% (rms) over a year.
Intermediate-time-scale stability of the IRAC photometry was 
measured by monitoring at least one secondary calibrator (near the ecliptic
plane) every 12 hr while IRAC is in nominal operations;
the intermediate-term photometry is stable with a 1\% dispersion (rms).
One of the secondary calibrators was found to have significantly
time-variable (5\%) mid-infrared emission,
with period (7.4 days) matching the optical light curve; it is possibly
a Cepheid variable.

\keywords{infrared, instrumentation, calibration}

\end{abstract}
                            
\section{Introduction}

The Infrared Array Camera (IRAC) was built at NASA Goddard Space Flight Center
under the direction of the Smithsonian Astrophysical Observatory
\citep{fazio}. It is the mid-infrared camera on the
Spitzer Space Telescope \citep{werner}, with four arrays 
observing at 3.6, 4.5, 5.8, and 8 $\mu$m. The absolute calibration
of the camera was performed in flight by comparing observed to
predicted brightness for a set
of stars that was selected and characterized before
launch \citep{mcisocal,cohenxiii}. This paper presents the in-flight
calibration results, including the observing strategy, the
predictions and measurements, and an assessment of the calibration
accuracy and stability of the instrument and the pipeline-processed
data provided by the Spitzer Science Center to observers.

\section{Calibrator Selection and Observing Strategy}

The primary calibrators (Table~\ref{primary1})
are a set of A main sequence and K giant stars.  
They were chosen from a larger list of candidates 
on the basis of having good spectral  and 
photometric data and no known evidence of variability \citep{stmisocal}.
The set of plausible calibrators 
was observed during the in-orbit checkout to select which
ones would become the primary calibrators for the nominal mission.
Table~\ref{calstar_ioc_table} lists the coordinates of the candidate primary calibrators 
that were observed during the in-orbit checkout, together with the IRAC
channels in which useful data were obtained.
Table~\ref{primary1} lists the primary calibrators that have been
observed during the nominal mission.

\begin{deluxetable}{rllclcrrrrr}
\tabletypesize{\scriptsize}
\rotate
\tablecaption{IRAC Primary absolute calibrators\label{primary1}}
\tablehead{\colhead{\#} & \colhead{Star\tablenotemark{a}} & \colhead{Other name} &\colhead{2MASSJ\tablenotemark{b}} 
&  \colhead{Type} & \colhead{$A_V$} &\colhead{Ks} 
&\colhead{[3.6]}&\colhead{[4.5]}&\colhead{[5.8]}&\colhead{[8.0]}}
\startdata
1	&NPM1p64.0581&HD 180609		&19124720+6410373	&A0V	     &0.387& 9.117&  9.123& 9.104& 9.107& 9.122  \\
2	&HD 165459		&		&18023073+5837381	&A1V	     &0.285& 6.584&  6.593& 6.575& 6.579& 6.591  \\
3	&NPM1p60.0581&BD+60 1753	&17245227+6025508	&A1V	     &0    & 9.645&  9.666& 9.653& 9.659& 9.669  \\
4	&1812095	&			&18120956+6329423	&A3V	     &0.378&11.596& 11.275&11.254&11.255&11.265  \\
5	&KF08T3	&			&17551622+6610116 &K0.5III   &0    &11.090& 10.944&10.986&10.985&10.943  \\
6	&KF09T1&GSC 04212-01074		&17592304+6602561 &K0III     &0    & 8.114&  8.046& 8.087& 8.086& 8.047  \\
7	&KF06T1	&			&17575849+6652293	&K1.5III   &0    &10.987& 10.762&10.906&10.833&10.793  \\
8	&KF06T2	&			&17583798+6646522	&K1.5III   &0.189&11.273& 11.065&11.193&11.133&11.095  \\
9	&NPM1p66.0578&GSC 04229-01455	&19253220+6647381	&K1III     &0    & 8.330&  8.249& 8.347& 8.310& 8.335  \\
10	&NPM1p67.0536&SAO 17718		&17585466+6747368 &K2III     &0.161& 6.409&  6.306& 6.428& 6.389& 6.348  \\
11	&NPM1p68.0422&BD+68 1022	&18471980+6816448	&K2III     &0.462& 6.816&  6.712& 6.822& 6.780& 6.742  \\
\enddata
\tablenotetext{a}{The star names used in this project are either abbreviated 
RA-only 2MASS designations, 
the designation NPM1 from the Lick Northern Proper Motion program \citep{klemola} with "p" replacing $+$, 
the designation KF from a north ecliptic pole optical/near-infrared survey \citep{kummel}.
These are not intended to replace the formal designations.}
\tablenotetext{b}{Designation of the corresponding 2MASS catalog entry, with `2MASSJ' prefix removed,
comprising the J2000 right ascension and declination {\it sans} punctuation (2 decimal places of 
seconds of right ascension and one decimal place of arcseconds of declination).}
\end{deluxetable}  

\begin{table}
\caption[]{IRAC calibrator candidates}\label{calstar_ioc_table} 
\begin{flushleft} 
\begin{tabular}{llcl} 
\hline
Star$^a$     &type	     & RA Dec (J2000) & Channels  \\ \hline\hline
     1732526 &     A3V &  17 32 52.6  +71 04 43.1 & 123 \\
     1739431 &     A3V &  17 39 43.2  +61 02 55.8 & 1234\\
     1740346 &     A5V &  17 40 34.7  +65 27 15.0 & 1234\\
     1743045 &     A5V &  17 43 04.5  +66 55 01.7 & 1234\\
     1757132 &     A3V &  17 57 13.2  +67 03 40.0 & 1234\\
     1802271 &     A3V &  18 02 27.2  +60 43 35.6 & 1234\\
     1805292 &     A1V &  18 05 29.3  +64 27 52.1 & 1234\\
     1808347 &     A3V &  18 08 34.8  +69 27 28.7 & 1234\\
     1812095 &     A2V &  18 12 09.6  +63 29 42.3 & 1234\\
     1812523 &     A1V &  18 12 52.4  +60 02 32.0 & 12 4\\
      KF01T3 &   K1III &  18 03 45.5  +66 56 03.0 & 1234\\
      KF01T5 &   K1III &  18 04 03.8  +66 55 43.0 & 1234\\
      KF02T1 &   K0III &  18 02 01.7  +66 37 39.0 & 1234\\
      KF02T3 &   K0III &  18 03 21.8  +66 34 08.0 & 1234\\
      KF03T1 &   K0III &  17 57 43.9  +66 26 55.0 & 1234\\
      KF03T2 & K1.5III &  17 57 51.4  +66 31 03.0 & 1234\\
      KF03T4 &   K1III &  17 59 03.9  +66 30 59.0 & 1234\\
      KF05T1 &   K1III &  17 59 12.1  +66 41 36.0 & 1234\\
      KF06T1 & K1.5III &  17 57 58.5  +66 52 29.0 & 1234\\
      KF06T2 & K1.5III &  17 58 37.9  +66 46 52.0 & 1234\\
      KF06T3 &   K1III &  17 58 50.2  +66 49 40.0 & 1234\\
      KF07T2 & K1.5III &  18 01 40.7  +65 59 44.0 & 1234\\
      KF07T3 & K1.5III &  18 02 14.8  +66 03 00.0 & 1234\\
      KF08T3 & K0.5III &  17 55 16.2  +66 10 11.0 & 1234\\
      KF09T1 &   K0III &  17 59 23.0  +66 02 56.0 & 1234\\
 NPM1p61.0569 & K0.5III &  17 23 25.9  +61 12 40.7 & 1234\\
 NPM1p61.0570 &   K1III &  17 26 11.6  +61 00 48.1 & 1234\\
 NPM1p61.0582 &   K1III &  17 36 55.6  +61 40 58.1 & 1234\\
 NPM1p64.0590 &   K0III &  19 25 14.9  +65 01 58.9 & 1234\\
 NPM1p66.0507 &   K2III &  17 31 22.1  +66 46 35.3 & 1234\\
 NPM1p66.0513 &   K2III &  17 43 53.7  +66 31 49.8 & 1234\\
 NPM1p66.0514 & K0.5III &  17 45 45.4  +66 48 45.4 & 1234\\
 NPM1p67.0536 &   K2III &  17 58 54.5  +67 47 37.4 & 1234\\
 NPM1p68.0412 &   K2III &  18 28 24.8  +68 58 51.5 & 1234\\
\hline
\end{tabular} 
\end{flushleft} 
$^a$ see note $a$ of Table~\ref{primary1}.
\end{table}

Primary calibrators are observed with all 4 IRAC arrays,
every instrument campaign.
The primary calibrators are located in the Spitzer continuous viewing zone,
a region approximately $10^\circ$ in radius centered on the north
ecliptic pole.
The south ecliptic pole is also continuously visible, but it 
saves observing time to have all calibrators close together
on the sky.
After the first six months of the nominal science mission, the radius of
the continuous viewing zone was decreased to 7\fdg5, making
two of the primary calibrators visible for only 8 months each
year.

Secondary calibrators are observed approximately every 12 hr while IRAC is on; they are
intended to monitor the gain stability during each observing campaign. Because
routine Spitzer operations require pointing the high-gain antenna at the Earth
every $\sim 12$ hr to downlink data, we chose a network of secondary calibrators that can
be rapidly observed after each downlink. The Spitzer orbit is close to the
ecliptic plane, and the telescope points opposite the high-gain antenna.  Thus
when the antenna points at Earth, the telescope points somewhere near the ecliptic
plane at a longitude that changes with a period of about one year. Secondary
calibrators spaced along the ecliptic can therefore be observed
efficiently before or after
telemetry downlinks, with little slewing overhead. The network of secondary
calibrators has two stars every $20^\circ$ of ecliptic longitude.  One or two
secondary calibrators are chosen for each campaign.

All calibration star observations are performed by placing the star
in 5 positions: the center of the array and the center of the four
quadrants of the array. A single frame is taken at each position.
Exposure times are set to place the peak
brightness approximately 1/3 of saturation, so that the signal-to-noise 
is high and the nonlinearity corrections are accurate. 
The IRAC arrays saturate at $\sim 10^5$ electrons or 40000 DN 
(whichever is lower).

\begin{deluxetable}{rlllllcccc}
\tabletypesize{\tiny}
\tablecaption{IRAC Secondary calibrators for each campaign\label{secondary}}
\tablehead{\colhead{\#} & \colhead{Campaign} & \colhead{Star} &\colhead{RA (J2000)} &  \colhead{Dec} & \colhead{Type} & \colhead{[3.6]} & \colhead{[4.5]} &\colhead{[5.8]}&\colhead{[8.0]} }
\startdata
 1 & IRAC003500& HD 218528   & 23 08 49.4 & $-$08 48 27 & K0III	  & 6.311$\pm$0.025 & 6.367$\pm$0.045& 6.361$\pm$0.024& 6.309$\pm$0.021\\
   &           & SA115 554   & 23 41 31.0 & $+$01 26 26 & K1.5III & 9.071$\pm$0.021 & 9.213$\pm$0.025& 9.154$\pm$0.018& 9.099$\pm$0.019\\
 2 & IRAC003600& HD 4182     & 00 44 25.4 & $+$09 57 09 & A2V	  & 8.990$\pm$0.021 & 8.981$\pm$0.021& 8.974$\pm$0.021& 8.980$\pm$0.021\\
   &           & SA115 554   & 23 41 31.0 & $+$01 26 26 & K1.5III & 9.071$\pm$0.021 & 9.213$\pm$0.025& 9.154$\pm$0.018& 9.099$\pm$0.019\\
 3 & IRAC003700& BP 20 417   & 02 32 14.4 & $+$20 49 09 & K1.5III & 7.777$\pm$0.022 & 8.058$\pm$0.025& 7.998$\pm$0.018& 7.944$\pm$0.018\\
 4 & IRAC003800& BP 14 549   & 03 18 20.3 & $+$15 25 41 & K2III	  & 7.104$\pm$0.021 & 7.234$\pm$0.027& 7.192$\pm$0.018& 7.138$\pm$0.019\\
 5 & IRAC003900& HD 37725    & 05 41 54.4 & $+$29 17 51 & A3V	  & 7.876$\pm$0.023 & 7.870$\pm$0.023& 7.864$\pm$0.023& 7.869$\pm$0.023\\
   &           & HD 244937   & 05 34 54.2 & $+$30 03 28 & A1V	  & 8.512$\pm$0.023 & 8.488$\pm$0.023& 8.472$\pm$0.023& 8.487$\pm$0.023\\
 6 & IRAC004000& HD 55728    & 07 14 47.6 & $+$16 35 50 & A2V	  & 9.465$\pm$0.019 & 9.464$\pm$0.019& 9.461$\pm$0.020& 9.463$\pm$0.020\\
   &           & HD 55677    & 07 14 31.3 & $+$13 51 37 & A4V	  & 9.189$\pm$0.020 & 9.187$\pm$0.02 & 9.184$\pm$0.020& 9.187$\pm$0.020\\
 7 & IRAC004100& AKS95 408   & 08 48 52.6 & $+$19 03 42 & K0III	  & 9.187$\pm$0.020 & 8.511$\pm$0.044& 8.505$\pm$0.023& 8.453$\pm$0.019\\
   &           & HD 77823    & 09 05 15.8 & $+$14 14 43 & K1.5III & 6.183$\pm$0.021 & 6.325$\pm$0.025& 6.266$\pm$0.018& 6.211$\pm$0.018\\
 8 & IRAC004200& HD 89562    & 10 20 08.9 & $+$03 11 56 & K0III	  & 6.949$\pm$0.025 & 7.004$\pm$0.045& 6.998$\pm$0.024& 6.946$\pm$0.021\\
   &           & HD 90593    & 10 27 44.2 & $+$04 28 20 & K0III	  & 7.049$\pm$0.026 & 7.105$\pm$0.045& 7.099$\pm$0.025& 7.047$\pm$0.023\\
 9 & IRAC004800& HD 99253    & 11 25 14.9 & $-$05 52 08 & K1.5III & 6.239$\pm$0.022 & 6.381$\pm$0.026& 6.322$\pm$0.020& 6.267$\pm$0.020\\
10 & IRAC004900& HD 109441   & 12 34 47.7 & $-$11 46 47 & K0III	  & 7.123$\pm$0.024 & 7.179$\pm$0.044& 7.173$\pm$0.024& 7.121$\pm$0.021\\
   &           & HD 113745   & 13 05 57.0 & $-$08 41 33 & A0V	  & 8.835$\pm$0.020 & 8.832$\pm$0.02 & 8.829$\pm$0.020& 8.834$\pm$0.020\\
11 & IRAC005000& HD 131769   & 14 55 51.2 & $-$12 12 28 & K1.5III & 6.552$\pm$0.022 & 6.682$\pm$0.026& 6.616$\pm$0.019& 6.567$\pm$0.020\\
   &           & HD 135248   & 15 14 21.4 & $-$08 50 39 & K1III	  & 6.596$\pm$0.025 & 6.704$\pm$0.032& 6.663$\pm$0.021& 6.611$\pm$0.022\\
12 & IRAC005100& HD 137429   & 15 26 29.7 & $-$14 28 35 & K0III   & 6.181$\pm$0.027 & 6.227$\pm$0.046& 6.217$\pm$0.026& 6.169$\pm$0.023\\
   &           & HD 145152   & 16 09 51.2 & $-$12 19 09 & K2III   & 6.570$\pm$0.024 & 6.705$\pm$0.029& 6.666$\pm$0.022& 6.609$\pm$0.022\\
13 & IRAC005200& HD 155334   & 17 12 01.2 & $-$25 40 54 & A1V     & 7.716$\pm$0.022 & 7.714$\pm$0.022& 7.710$\pm$0.023& 7.714$\pm$0.023\\
   &           & HD 156896   & 17 20 55.5 & $-$20 21 35 & A0V     & 9.364$\pm$0.021 & 9.355$\pm$0.021& 9.349$\pm$0.021& 9.357$\pm$0.021\\
14 & IRAC005300& HD 184837   & 19 37 16.4 & $-$24 40 11 & A0V     & 9.595$\pm$0.021 & 9.587$\pm$0.021& 9.581$\pm$0.021& 9.588$\pm$0.021\\
15 & IRAC005400& HD 195061   & 20 29 33.2 & $-$18 13 12 & A0V     & 9.330$\pm$0.022 & 9.318$\pm$0.022& 9.310$\pm$0.022& 9.319$\pm$0.022\\
\enddata
\end{deluxetable}  

\section{Photometry\label{photometrysection}}

As input, we use the ``basic calibrated data'' (BCD) 
generated by the IRAC science pipeline (version S10
\footnote{See the IRAC Data Handbook and pipeline history log 
and Pipeline Description Document
at {\tt http://ssc.spitzer.caltech.edu/irac/data.html} for details.}). 
In summary, the pipeline removes the electronic bias, subtracts a dark sky 
image generated
from observations of relatively empty sky near the ecliptic pole, 
flat-fields using a super-flat generated from the first year's
calibration observations of relatively blank fields near the ecliptic
plane, and linearizes
using laboratory measurements of each pixel's response to a
calibration lamp in frames of varying length.
For each BCD image in each standard star observation, aperture photometry
was used to determine the source flux. The target is located as the 
image maximum after spatial filtering (low-pass, median, 7 pixel width) 
to reduce noise and cosmic rays; this filtered image was used {\it only}
for source identification. 
(Some of the candidate calibrators observed during in-orbit checkout
have other stars nearby; they were located using the coordinates of
the star and the astrometric calibration of the images.)
The target location is then refined
using a centroiding algorithm ({\tt cntrd} in IDLphot).
We place the image into electron units for proper error
estimation; from the pipeline-processed
images (which are in MJy~sr$^{-1}$), we multiply by the gain 
({\tt GAIN} in the header; elec/DN) and
exposure time ({\tt EXPTIME} in the header)
and divide by the calibration factor ({\tt FLUXCONV}
in the header; MJy~sr$^{-1}$~DN$^{-1}$~s) that had been
used in the pipeline; the scaled image at pixel $i,j$ 
is then $I_{ij}$ in electrons.
To get the absolute brightness of each pixel, we also added the brightness
of the zodiacal light at the time of the sky-dark observation
(approximately 0.036, 0.18, 1.6, 4.4 MJy~sr$^{-1}$ in channels 
1, 2, 3, and 4, respectively); this is
only important for error propagation as it is a constant over the image.
The background is calculated by a robust average in an annulus spanning 
12 to 20 pixel radius centered on the target
\begin{equation}
I_{sky} = \Sigma_{sky}^{robust} { I_{ij} } / N_{sky}
\end{equation}
The on-source flux is
calculated by summing over a 10-pixel radius centered on the target
{\tt aper} in IDLphot). 
\begin{equation}
F_{elec} = \Sigma_{on}{ I_{ij} - I_{sky} }.
\end{equation}
The electron production rate, $N_e = F_{elec}/t$ 
is proportional to the stellar flux.
\def\extra{
The flux in physical units can be calculated by 
dividing by the gain and exposure time, multiplying by the
the calibration factor, and multiplying by the pixel solid
angle:
\begin{equation}
F_{Jy} = F_{elec} \frac{{\cal C}\Omega_{pix}}{G t}.
\end{equation}
}

An array-location-dependent photometric correction must be applied to 
IRAC photometry, to account for the variation in pixel solid angle
(due to distortion) and the variation of the spectral response
(due to the tilted filters and wide field of view) over the array
\citep{horafov}. The photometric corrections are defined to be unity
in the center of the array, so their role in the data analysis here
is to remove a systematic error that would make the calibration
star stars, which were observed at 5 widely-separated positions on the array
(see Table~\ref{delta_flat} for the exact locations),
have dispersions significantly higher than the measurement uncertainties.
Table~\ref{delta_flat} shows the percentage corrections that were applied;
specifically, if the table entries are $f_{ij}$ then the fluxes in 
channel $i$ at position $j$ were divided by $1+f_{ij}/100$.

\begin{table}
\caption[]{Array-location-dependent Photometric Corrections (\%)\label{delta_flat} }
\begin{flushleft} 
\begin{tabular}{cccrrrr} 
\hline
& & & \multicolumn{4}{c}{Channel}\\ 
index & column & row & 1 & 2 & 3 & 4 \\ \hline
0 & 128 & 128 & 0.00 & 0.00 &  0.00 & 0.00 \\
1 & 64 & 192  & 0.52 & 0.53 &  3.86 & -1.31 \\
2 & 192 & 192 & 0.46 & 1.75 & -0.85 &  1.50\\
3 & 192 & 64  & 1.34 & 1.56 & -1.63 &  2.66\\
4 & 64 & 64   & 1.34 & 0.68 &  3.39 & -0.39\\
\hline
\end{tabular} 
\end{flushleft} 
\end{table}  

A photometric correction must also be applied to IRAC photometry to account for 
the variation in the flux of a source
as a function of location within a pixel. Sources placed directly in the center
of a pixel give higher count rates than those placed near a pixel edge.
The effect may be due to a nonuniform quantum
efficiency across each pixel. We refer to the `pixel phase' as the distance of
the centroid of a star from the center of the pixel containing the centroid.
The pixel-phase-corrections are normalized such that randomly-placed sources will
have equal chances of requiring an increase or decrease in the flux.
For randomly placed sources this location is $1/\sqrt{2\pi}$ pixels from the center. 
The pixel phase dependence is
only clearly detected in channel 1, and an approximate fit is
\begin{equation}
f_{phase}=1+0.0535 (1/\sqrt{2\pi} - p)
\label{pphase}
\end{equation}
where $p$ is the distance (in pixels)
from the source centroid to
the center of the pixel containing that centroid. 
Observed fluxes in channel 1 were divided by $f_{phase}$.
The channel 1 calibration stars tended
on average to fall closer to the centers of their pixels than 
a random distribution, 
and the median correction $\langle f_{phase}\rangle =1.0$\%.

The photometric uncertainty estimate includes three terms.
First, the uncertainty in the subtracted
sky background is the rms of the pixel values in the 
sky annulus, $\sigma_{sky}$, divided by the 
square-root of the number of pixels in the sky annulus
and multiplied by the number of on-target aperture
pixels:
\begin{equation}
\sigma_{sky} = \Re_{sky} N_{on}/\sqrt{N_{sky}} ,  
\end{equation}
where $\Re_{sky}$ denotes the root-mean-square (rms) intensity in
the sky annulus.
Second, noise due to sky variations within the on-source aperture is
\begin{equation}
\sigma_{sky,on} = \Re_{sky} \sqrt{ N_{on} } .              
\end{equation}
And third, the Poisson noise on the total background-subtracted
counts within the on-source aperture is
\begin{equation}
\sigma_{poisson} = \sqrt{ F_{elec} }.                       
\end{equation}
The uncertainty in the flux is then the quadrature sum of the
three error terms:
\begin{equation}
\sigma_F = \sqrt{\sigma_{sky}^2 + \sigma_{sky,on}^2 + \sigma_{poisson}^2}.
\label{sigmaf}
\end{equation}
We have not explicitly included random variations in the instrument
or its calibration; instead we attempt to measure those variations
through the dispersion in photometry on a large number of images.
All three terms in equation~\ref{sigmaf} contribute significantly to the
photometric uncertainty. For example, for NPM1p64.0581, the proportion
$\sigma_{sky}$:$\sigma_{sky,on}$:$\sigma_{poission}$ is
0.64:0.66:0.39 for channel 1, 0.71:0.54:0.45 for channel 2,
0.69:0.58:0.43 for channel 3, and 0.80:0.33,0.50 for channel 4.
The differences in proportion are due to 
background brightness and dispersion, source flux, and exposure time
in each channel. The $\sigma_{sky}$ can be reduced somewhat
by increasing the size of the sky annulus.
Both $\sigma_{sky,on}$ and $\sigma_{sky}$ can be reduced by decreasing the size
of the on-source aperture, but then we would lose some flux and require
larger aperture corrections. The choice of on and sky apertures used
here is optimized for calibration purposes, where we attempt to gather
as much of the star's flux as possible. For source detection,
a smaller on-source aperture is recommended. All of the calibration stars
are bright; for example, for NPM1p64.0581, the flux-to-uncertainty
ratio, $F/\sigma_F$ is
158, 106, 123, and 96 for channels 1, 2, 3, and 4, respectively,
in a single BCD image.

The flux and photometric uncertainty were calculated for each image of
each calibration star; 15,341 fluxes were measured for IRAC campaigns 1-15.

\section{Photometric stability}

\begin{figure}[th]
\epsscale{1}
\plotone{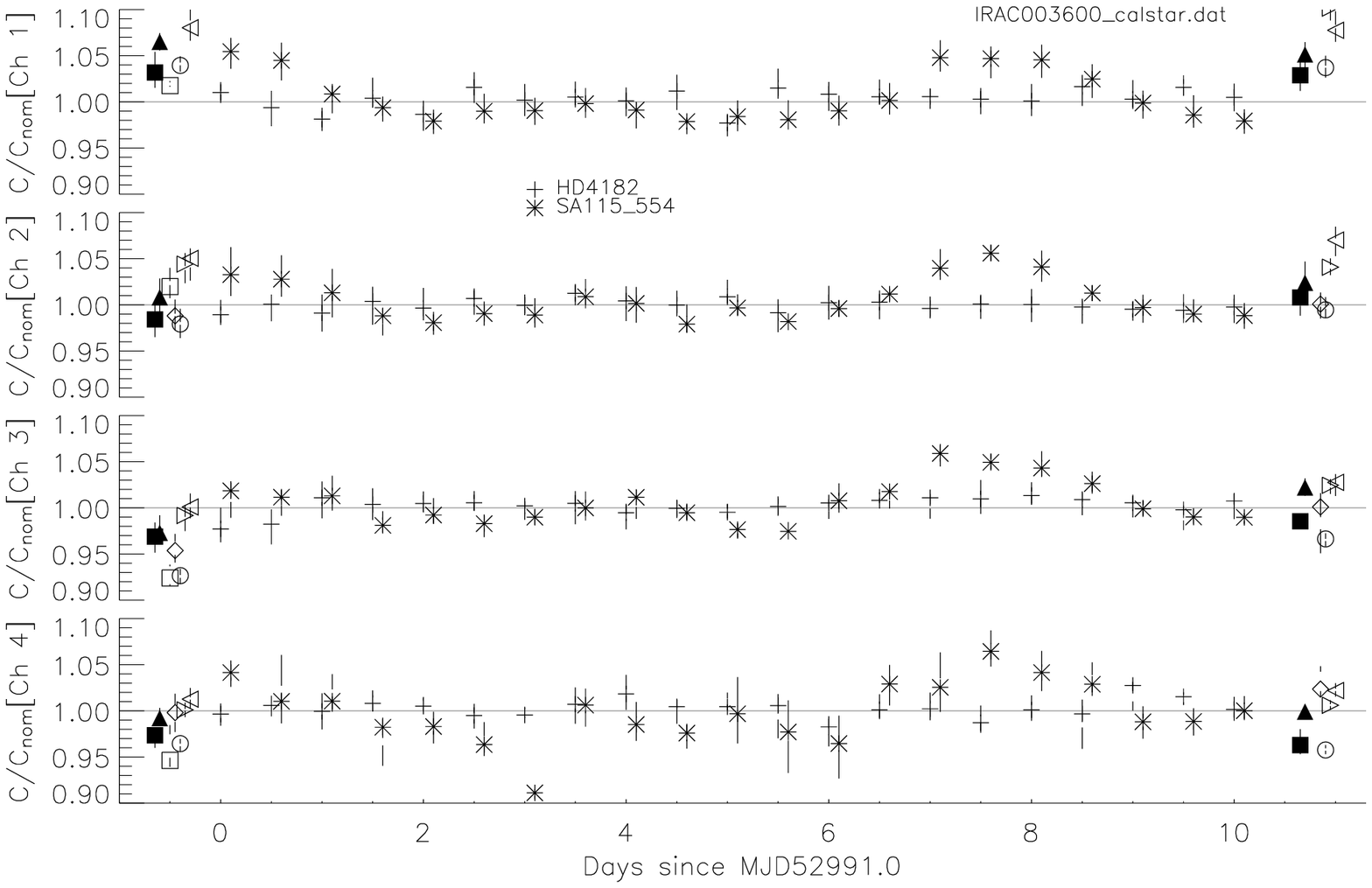}
\epsscale{1}
\figcaption[f1.eps]{Summary of calibration star results during 
the second IRAC campaign of the routine science mission. 
Each point is a measurement of the absolute calibration factor,
normalized by the new calibration factors derived in this paper
[0.1104, 0.1390, 0.6024, 0.2083  (MJy/sr)/(DN/s)].
The clusters of points and the beginning of each plot are the
primary calibrators, which always have the same symbol type in each plot.
Filled symbols are K giants and open symbols are A dwarfs. The primary 
calibrators (and their symbols) are NPM1p64.0581 (filled square),
HD 165459 (filled triangle pointing up), NPM1p60.0581 (filled diamond),
1812095 (filled triangle pointing down), KF08T3 (open square),
KF09T1 (open triangle pointing up), KF06T1 (open diamond), 
KF06T2  (open triangle pointing down), NPM1p66.0578 (open circle),
NPM1p67.0536 (open triangle pointing right), and
NPM1p68.0422 (open triangle pointing left). Two error bars are shown for
each point, indicating the rms dispersion and the uncertainty in the
mean. Spread throughout each campaign are the secondary calibrators, observed
every 12 hr. One or more secondary calibrators are observed each campaign,
with source names indicated in a legend for each campaign.
One of the two secondary calibrators (SA115-554) is
apparently a variable star and has been removed from our calibrator list.
\label{plot_calstar_campaign}}
\end{figure}

To monitor the stability of the camera on various timescales, the 
primary and secondary calibrators were all reduced uniformly and 
plotted together. Figure~\ref{plot_calstar_campaign} shows the results
for the second IRAC nominal operations campaign. 

\begin{figure}[th]
\epsscale{1}
\plotone{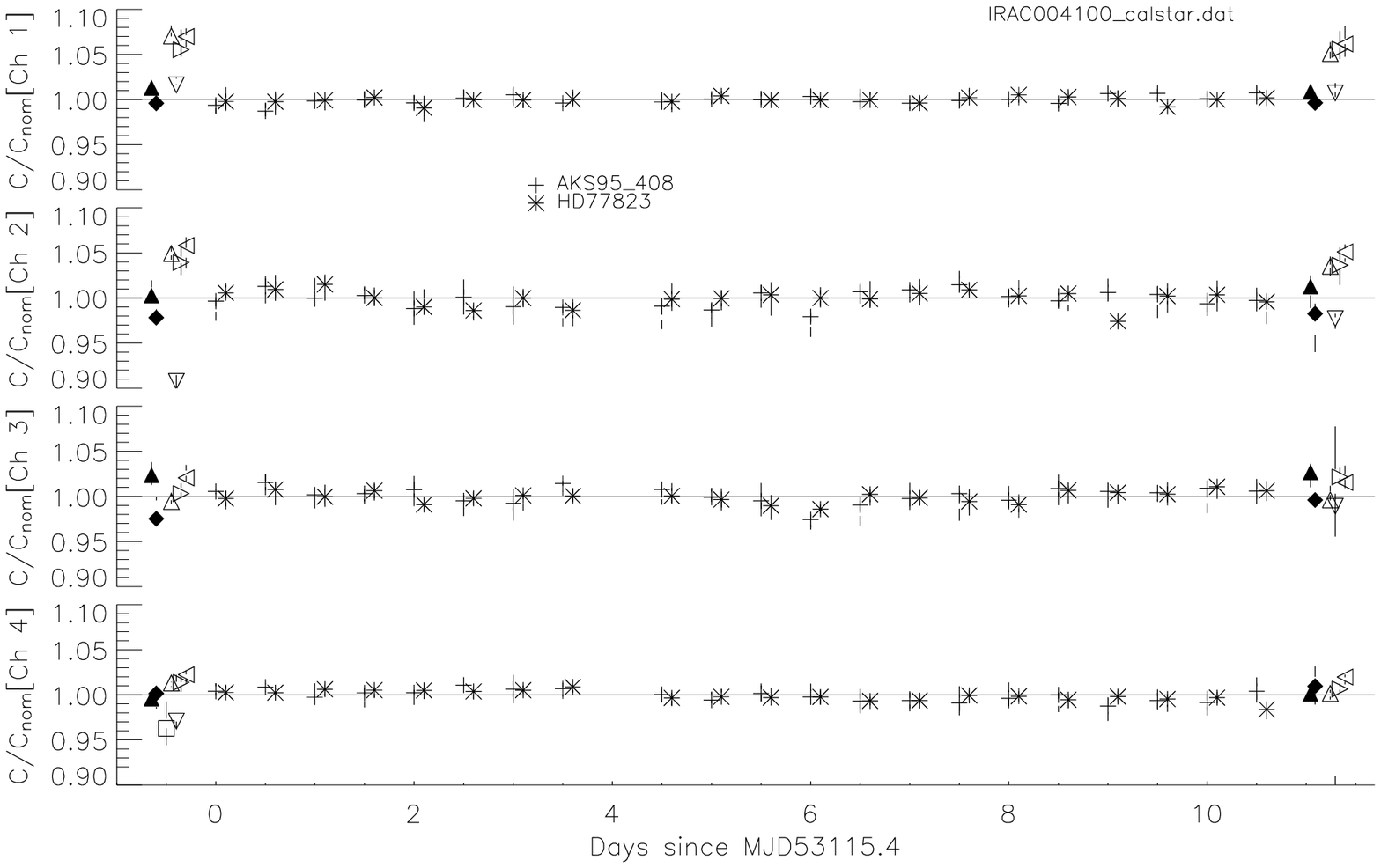}
\epsscale{1}
\figcaption[f2.ps]{Same as Figure~\ref{plot_calstar_campaign}, but
for campaign IRAC004100.}
\end{figure}

\begin{figure}[th]
\epsscale{1}
\plotone{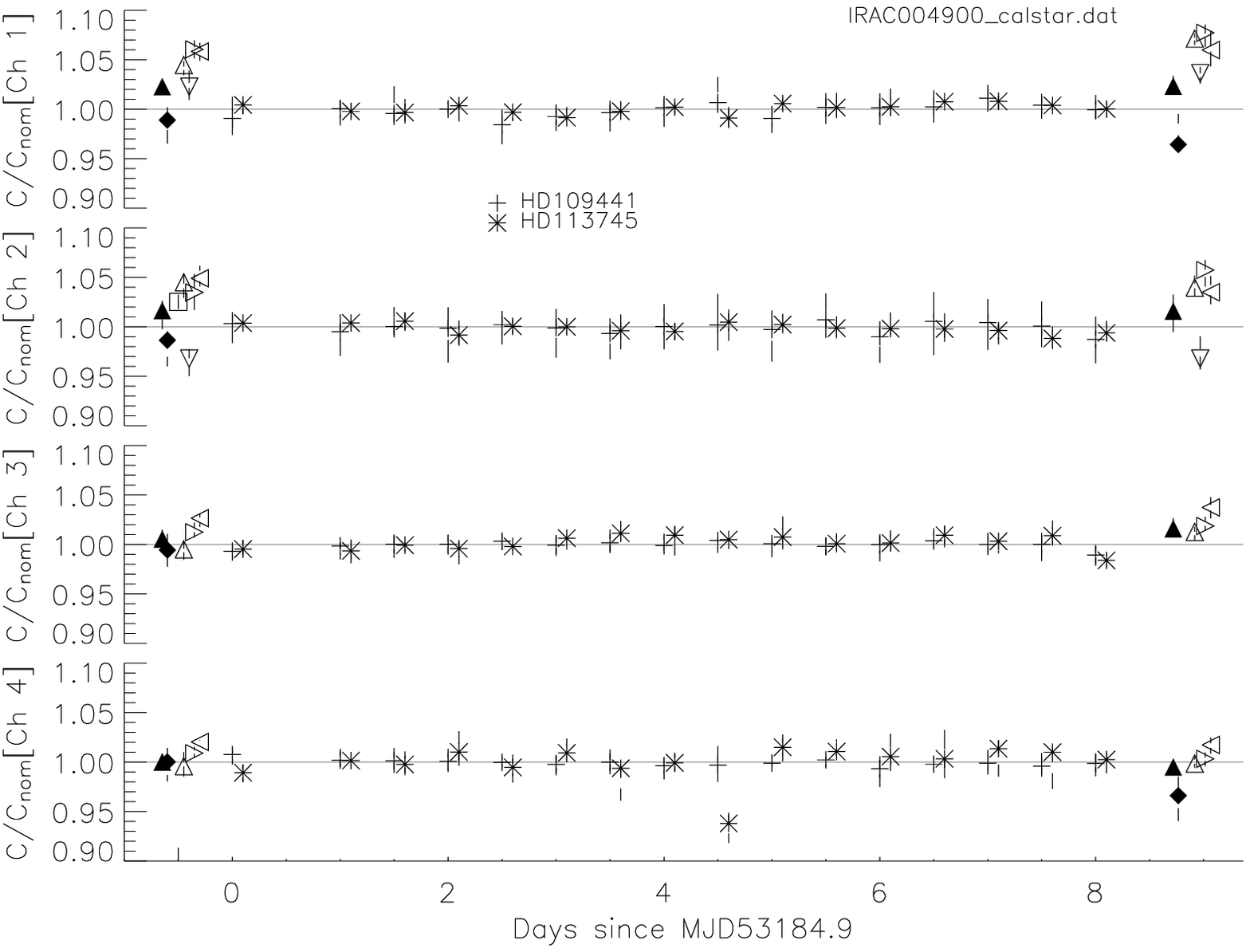}
\epsscale{1}
\figcaption[f3.eps]{Same as Figure~\ref{plot_calstar_campaign}, but
for campaign IRAC004900. }
\end{figure}

\begin{figure}[th]
\epsscale{1}
\plotone{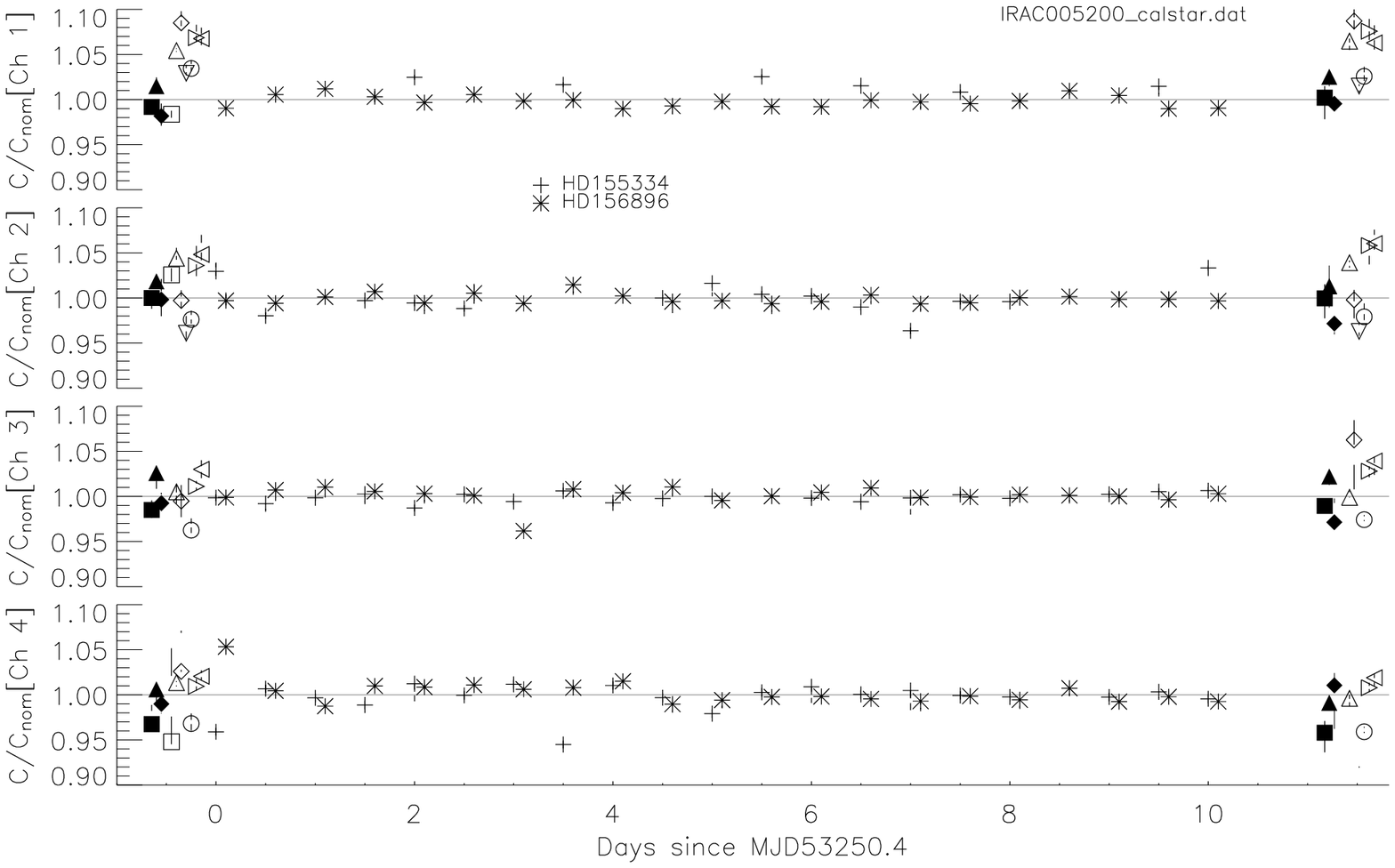}
\epsscale{1}
\figcaption[f4.eps]{Same as Figure~\ref{plot_calstar_campaign}, but
for campaign IRAC005200. }
\end{figure}

\begin{figure}[th]
\epsscale{1}
\plotone{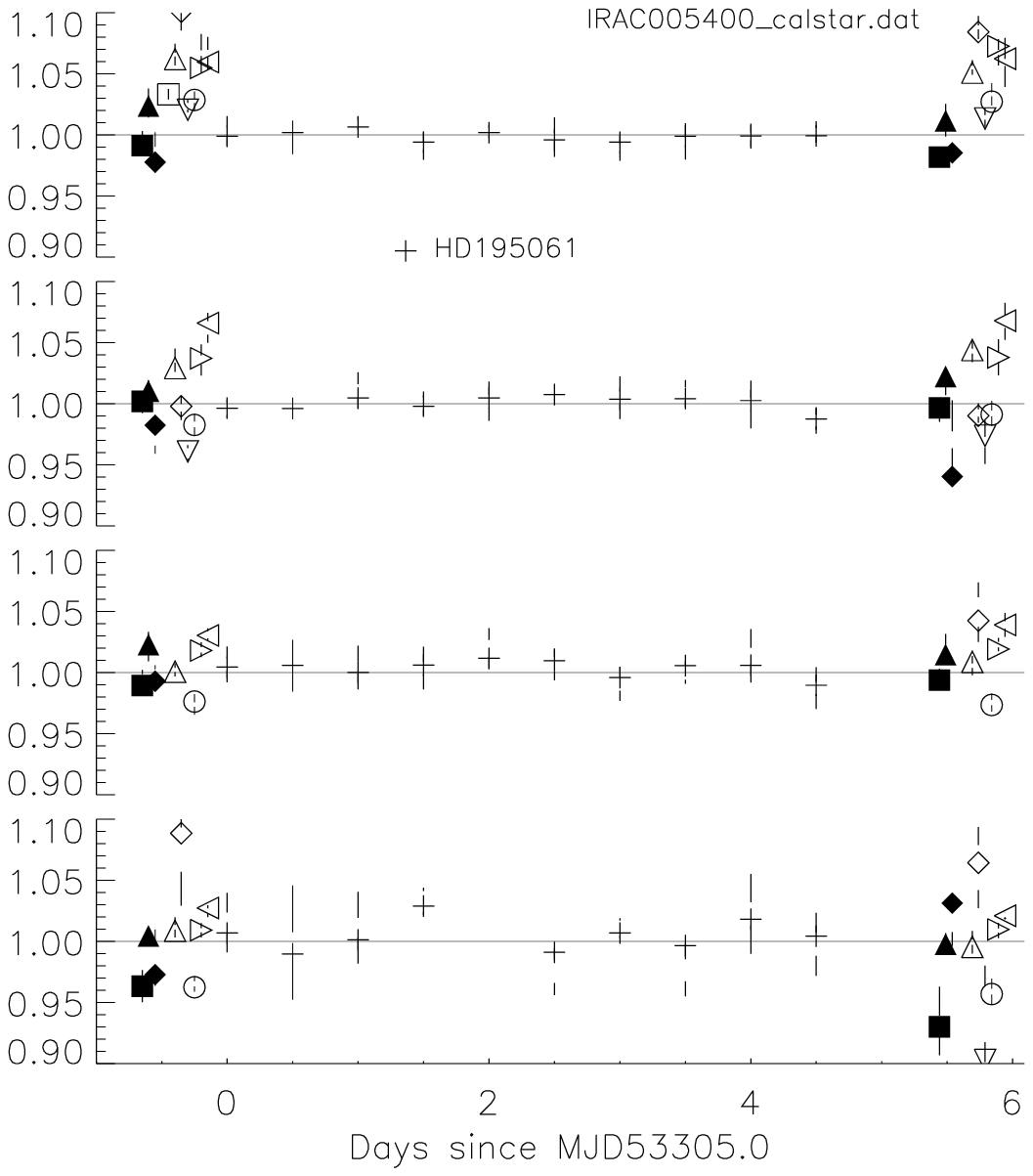}
\epsscale{1}
\figcaption[f5.eps]{Same as Figure~\ref{plot_calstar_campaign}, but
for campaign IRAC005400. This campaign length is typical of the second year
of routine operations.}
\end{figure}

\subsection{Stability on weekly timescales}

The relative repeatability of IRAC was measured by taking the root-mean-square 
variations of each
secondary calibration star photometry during a campaign. The distribution of
these variances among all of the campaigns has a $\chi^2$ probability distribution
function. The median of these rms values is 1.7\%, 2.2\%, 1.5\%, and 1.6\%
in channels 1, 2, 3, and 4, respectively. The variance of the rms is 
consistent with the median photometric measurement error, 
which is 0.5\%, 0.9\%, 0.7\%, and 0.9\% in channels 1, 2, 3, and 4, respectively.
Some stars have higher dispersions, but only one star at one wavelength (compared to
a total of 20 stars at 4 wavelengths) has an rms dispersion larger than 5\%. This
is a measure of the repeatability of photometry within a campaign, which can be
summarized as being stable to 2\%.

\subsection{Stability on long timescales}

The camera is evidently stable on all timescales that have been checked.
Table~\ref{longtermtab} summarizes the statistics characterizing
the long-term stability for 8 primary calibrators over 11 months.
To assess the stability, we use the statistics for individual calibrators in
Table~\ref{longtermtab}. Normalizing the count rates
for each primary calibrator by their median, the long-term dispersion in the
photometry becomes 1.7\%, 0.9\%, 0.9\%, and 0.5\% in channels
1, 2, 3, and 4, respectively. 

\begin{figure}[th]
\epsscale{1}
\plottwo{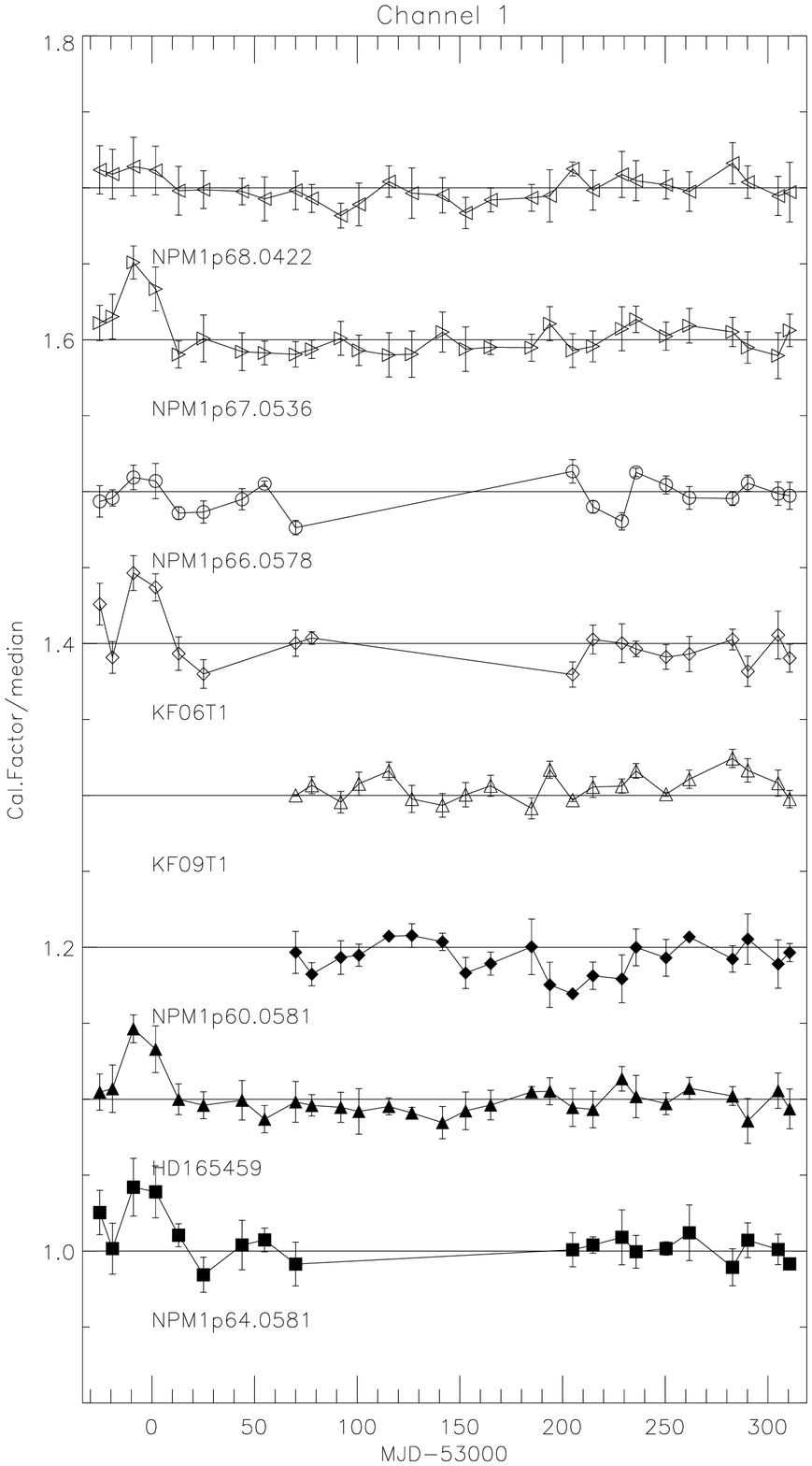}{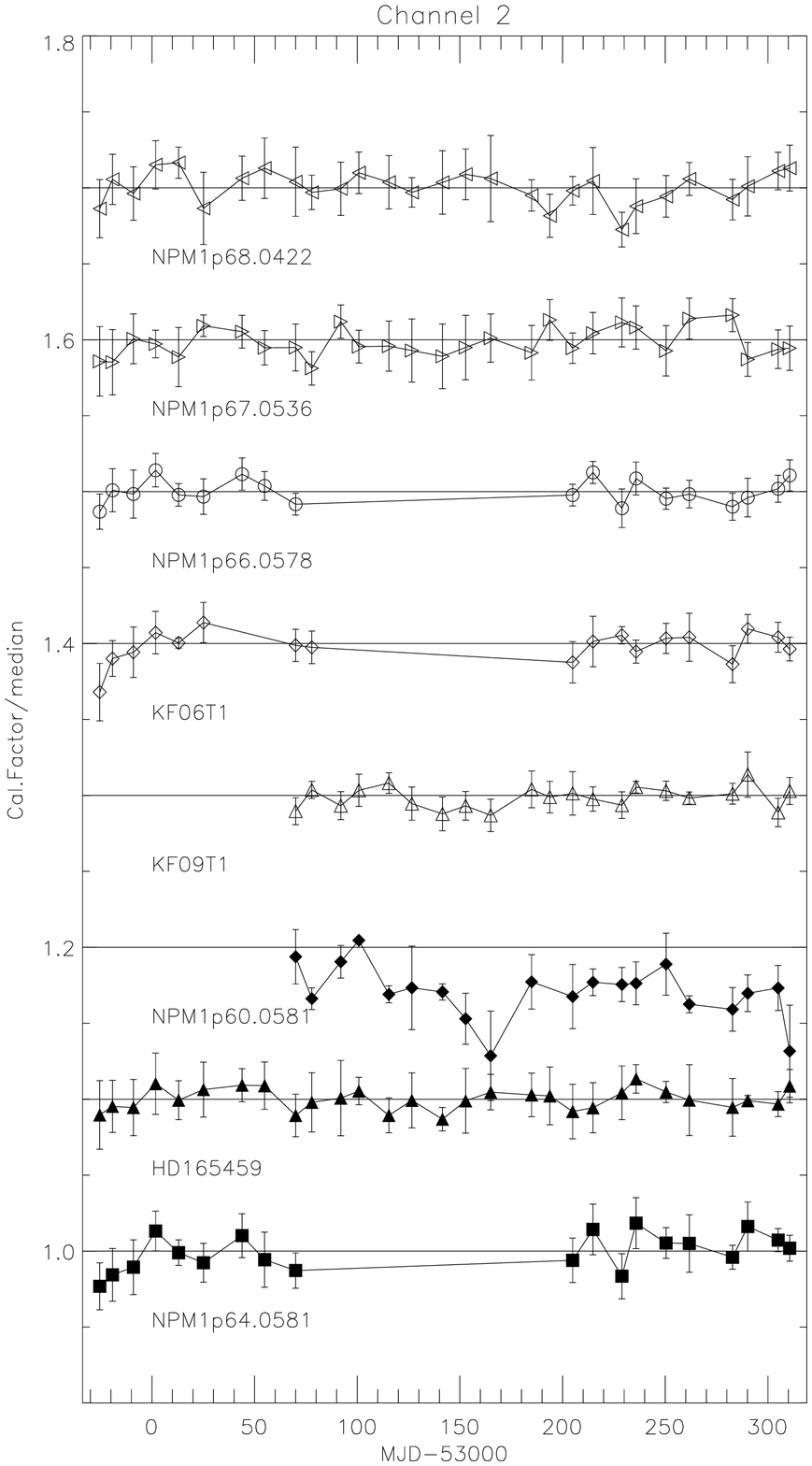}
\epsscale{1}
\figcaption[f6a.eps]{Long-term monitoring of the calibration factor
derived from individual primary calibrators. Each panel is for a different
IRAC channel: 1 (top left), 2 (top right), 3 (bottom left), and 4 
(bottom right). Within each panel, the calibration factors
for each primary calibrator have been normalized to their median,
and they have been offset from the previous curve by 0.1.
\label{longterm}}
\end{figure}

To determine whether the photometry has significant long-term
variations (due to source or instrument), 
we inspected the long-term trends for each calibrator.
Figure~\ref{longterm} shows the trend of the count rates
over the first 11 months of the mission, and the last two columns
in Table~\ref{longtermtab} show least-squares-fitted slopes to
the temporal variations. An instrumental trend will be the same in
the for each primary calibrator, while stellar variability will appear 
only in the individual star. Measurement
errors generate dispersion (both random and systematic) that should not
correlate between stars. 
Only one (KF09T1) of the 32 primary trend plots (8 calibrators, 4 channels)
shows a slope that is formally significant.
\footnote{
This example shows how a simple monitoring program, with
observations of 5 frames each, every 10--20 days, can limit
variability at the 2\% level. Such a monitoring program takes approximately
2 hr of observing time per year, including slews to the target
and observing at all 4 wavelengths.
}
The other primary
calibrators are all stable at a level of $< 1.5$\%/yr. 

\def\extra{
Channel 1 shows an apparent instrumental trend, with the inferred
calibration factor decreasing with time.
This could be due to an increase in the detector response versus time,
a change in the normalization of the flat-field, or a dependence of
the detector response on the annealing strategy.
The temporal variation in channel 1 does not appear to be a
linear or even a smooth long-term trend, and it is better described by a step 
function, with the first two campaigns constant at one value and the remainder
constant at another value. For the third through eleventh campaigns,
the slope is insignificant ($+0.5 \pm 0.3$\%/yr).
The calibration difference between the first two campaigns and
the third and subsequent campaigns is 3.9\%.
The flat-field used in pipeline processing for the first and (part of)
the second IRAC campaigns was developed independently from the
other flat-fields; no acceptable, routine flat-fields were possible
from the in-orbit checkout or the first campaign due to severe
long-term latent images from bright stars. These long-term latent
images are routinely removed by annealing, and the flat-fields
and darks are routinely scheduled close to anneals (to prevent their
corruption). This annealing and calibration-scheduling strategy
was not in place for the first two campaigns. 

FINISH THIS: WAS IT DUE TO THE FLAT---- YES!!! NOMOPS1 flat different by 
at least 2\% due to latents
}

\begin{table}
{\scriptsize
\caption[]{Long term (1 year) stability of IRAC primary calibration stars}\label{longtermtab} 
\begin{flushleft} 
\begin{tabular}{llllllll} 
\hline
  Star & Channel & Median$^a$ & RMS$^b$ & WTmean$^c$ & WTsig$^d$   &  slope$^e$    & sigslope$^e$\\ \hline\hline
  NPM1p64.0581 & 1 &  0.9672 &  0.0163 &  0.9709 &  0.0024 &  -0.01246 &  0.006787\\
  NPM1p64.0581 & 2 &  1.0105 &  0.0099 &  1.0071 &  0.0029 &   0.00316 &  0.004725\\
  NPM1p64.0581 & 3 &  1.0080 &  0.0082 &  0.9970 &  0.0012 &   0.00651 &  0.003850\\
  NPM1p64.0581 & 4 &  1.0017 &  0.0125 &  1.0081 &  0.0024 &   0.00142 &  0.005372\\
    HD165459 & 1 &  0.9892 &  0.0192 &  0.9815 &  0.0004 &  -0.01192 &  0.008299\\
    HD165459 & 2 &  1.0269 &  0.0067 &  1.0253 &  0.0020 &  -0.00212 &  0.002973\\
    HD165459 & 3 &  1.0412 &  0.0083 &  1.0422 &  0.0008 &   0.00366 &  0.003180\\
    HD165459 & 4 &  1.0390 &  0.0053 &  1.0373 &  0.0012 &   0.00127 &  0.002308\\
  NPM1p60.0581 & 1 &  0.9513 &  0.0124 &  0.9497 &  0.0018 &  -0.00802 &  0.007658\\
  NPM1p60.0581 & 2 &  0.9931 &  0.0183 &  1.0205 &  0.0008 &  -0.01825 &  0.010200\\
  NPM1p60.0581 & 3 &  1.0018 &  0.0828 &  0.9933 &  0.0020 &  -0.01570 &  0.010688\\
  NPM1p60.0581 & 4 &  1.0295 &  0.0178 &  1.0208 &  0.0020 &   0.00867 &  0.011478\\
      KF09T1 & 1 &  1.0268 &  0.0082 &  1.0292 &  0.0010 &   0.00806 &  0.005187\\
      KF09T1 & 2 &  1.0510 &  0.0073 &  1.0521 &  0.0017 &   0.00285 &  0.004707\\ 
      KF09T1 & 3 &  1.0195 &  0.0069 &  1.0178 &  0.0011 &  {\it 0.01244} &  {\it 0.003559}\\
      KF09T1 & 4 &  1.0424 &  0.0079 &  1.0470 &  0.0011 &   0.00365 &  0.005214\\
      KF06T1 & 1 &  1.0621 &  0.0241 &  1.0736 &  0.0019 &  -0.02734 &  0.011175\\
      KF06T1 & 2 &  1.0081 &  0.0083 &  1.0086 &  0.0018 &  -0.00551 &  0.003902\\
      KF06T1 & 3 &  1.0174 &  0.0643 &  1.0072 &  0.0055 &   0.04192 &  0.024166\\
      KF06T1 & 4 &  1.0648 &  0.0323 &  1.0745 &  0.0048 &   0.02743 &  0.014229\\
  NPM1p66.0578 & 1 &  1.0004 &  0.0143 &  0.9971 &  0.0004 &  -0.00989 &  0.006577\\
  NPM1p66.0578 & 2 &  0.9934 &  0.0096 &  0.9993 &  0.0019 &  -0.00448 &  0.004144\\
  NPM1p66.0578 & 3 &  0.9869 &  0.0081 &  0.9853 &  0.0013 &   0.00651 &  0.003476\\
  NPM1p66.0578 & 4 &  0.9986 &  0.0043 &  1.0010 &  0.0008 &  -0.00018 &  0.002175\\
  NPM1p67.0536 & 1 &  1.0351 &  0.0176 &  1.0295 &  0.0014 &   0.00070 &  0.007963\\
  NPM1p67.0536 & 2 &  1.0507 &  0.0097 &  1.0558 &  0.0025 &   0.00085 &  0.004416\\
  NPM1p67.0536 & 3 &  1.0383 &  0.0092 &  1.0374 &  0.0014 &   0.00062 &  0.004163\\
  NPM1p67.0536 & 4 &  1.0470 &  0.0044 &  1.0463 &  0.0010 &   0.00253 &  0.001915\\
  NPM1p68.0422 & 1 &  1.0349 &  0.0170 &  1.0336 &  0.0014 &  -0.01374 &  0.007078\\
  NPM1p68.0422 & 2 &  1.0697 &  0.0122 &  1.0667 &  0.0027 &  -0.00912 &  0.004984\\
  NPM1p68.0422 & 3 &  1.0497 &  0.0077 &  1.0496 &  0.0017 &   0.00767 &  0.003115\\
  NPM1p68.0422 & 4 &  1.0605 &  0.0057 &  1.0597 &  0.0010 &   0.00180 &  0.002322\\
  \hline
\end{tabular} 
\end{flushleft} 
}
$^a$ Median of calibration factors measured in each of the first 11 campaigns, 
     normalized to nominal values 0.1104, 0.1390, 0.6024, 0.2083  (MJy/sr)/(DN/s)
     for channels 1, 2, 3, and 4, respectively.
$^b$ Root-mean-square of normalized calibration factors.
$^c$ Weighted mean of normalized calibration factors; with the weight for each
     point being the root-mean-square dispersion of measurements during a campaign
$^d$ Formal statistical uncertainty in the weighted mean of the normalized
     calibration factors.
$^e$ Least-squares-fitted slope (and its statistical uncertainty) for the
     normalized calibration factors, in units of fractional variation per year.
     Apparently statistically significant slopes are in {\it italics}.
\end{table}  

\begin{figure*}[bh]
\plottwo{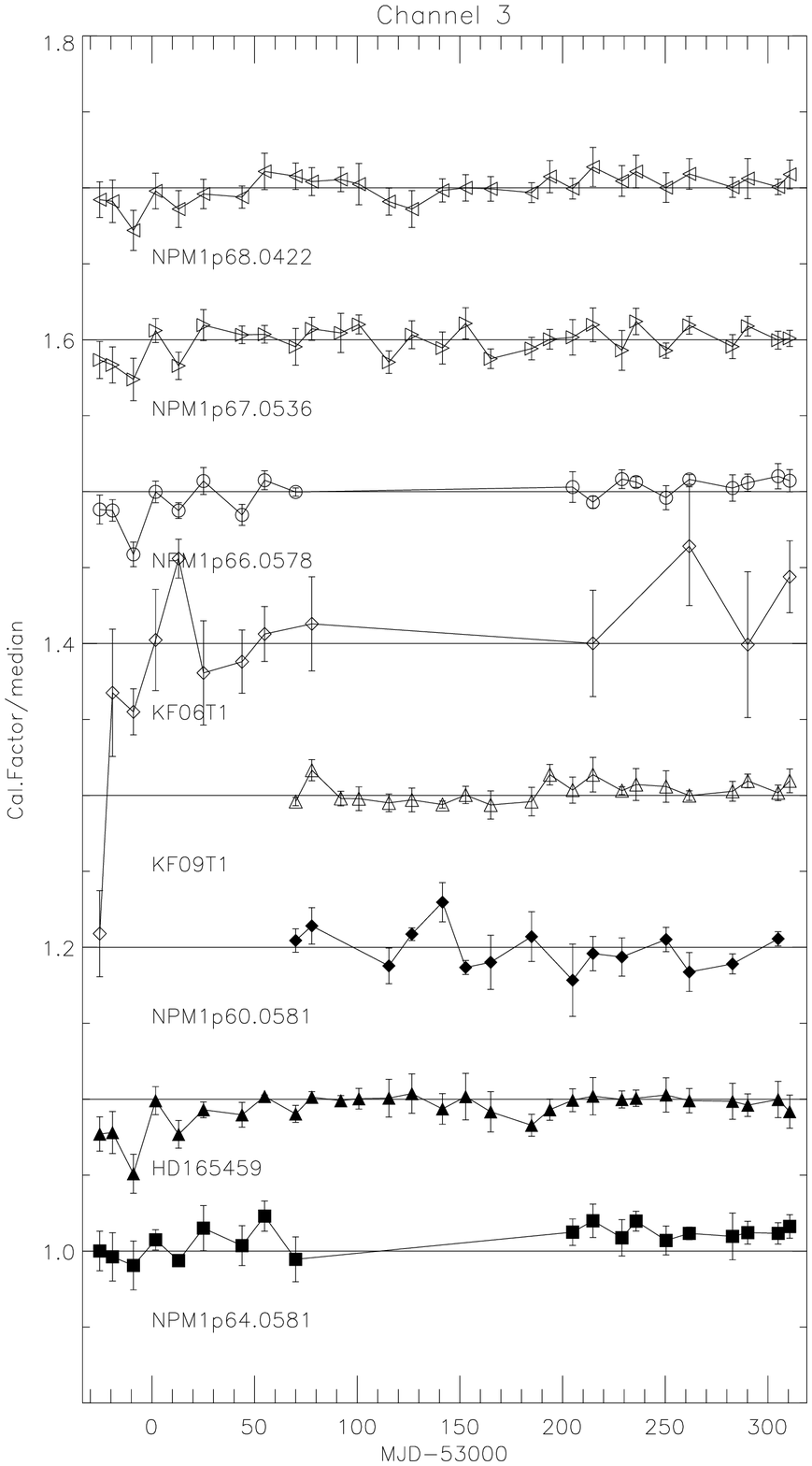}{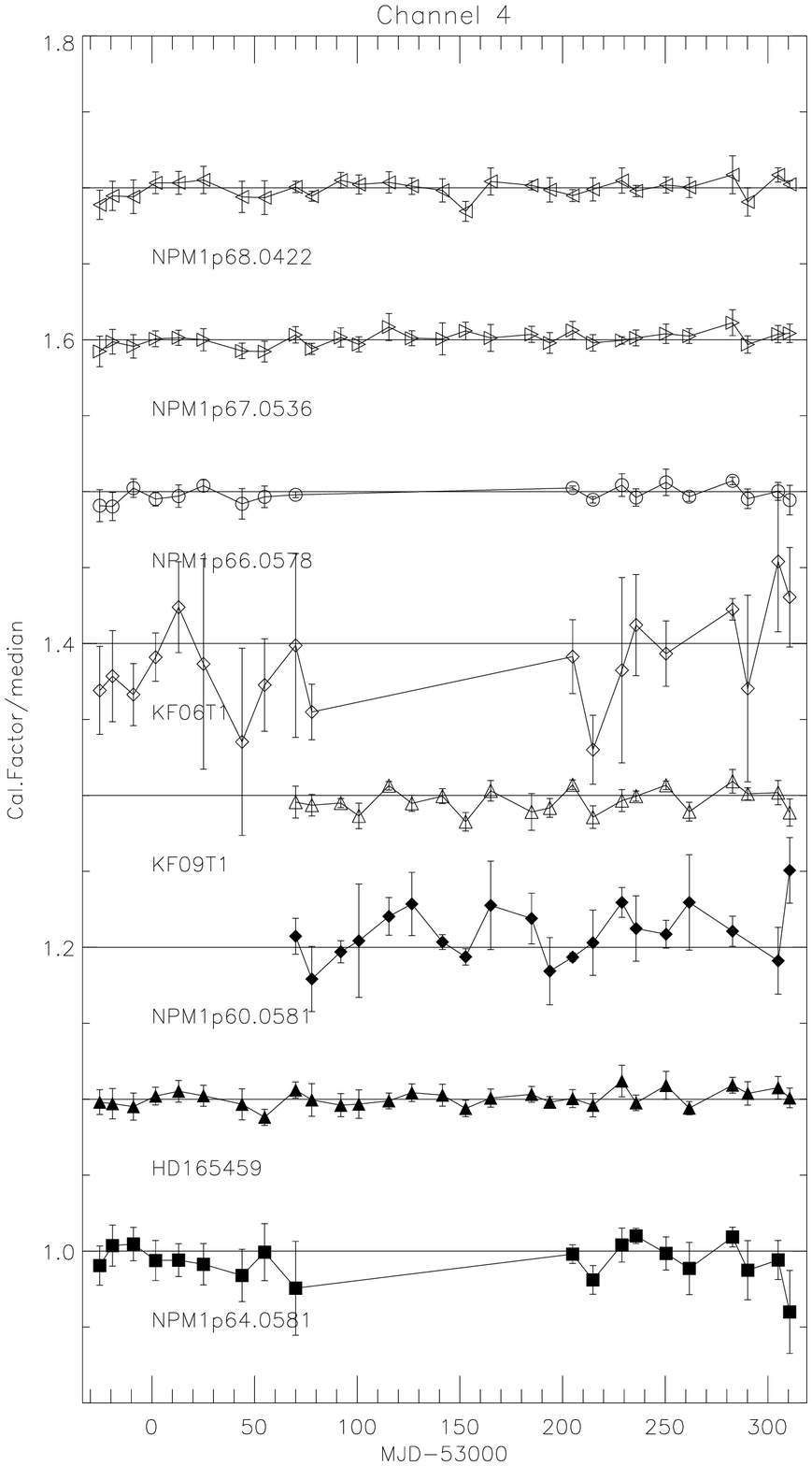}
\end{figure*}

\section{Absolute calibration}

\subsection{Flux prediction for primary calibrators}

For each primary calibrator, we predicted the count rate in each
IRAC channel.  The procedure was explained in detail by \citet{cohenxiii}.
In brief, the spectral type of each star was determined from ground-based visible
spectroscopy \citep{stmisocal}.  The intrinsic spectral energy distribution
for that type was determined from models for the A stars
\citep{kurucz}, or from empirical spectral templates derived from 
observations for the K giants \citep{cohenvii}.
For each star, an absolute normalization and the extinction
$A_{V}$ were derived by fitting the intrinsic SED to optical (UBV,
Hipparcos $H_p$, Tycho-2 $B_T$,$V_T$), near infrared (2MASS), and mid-infrared
(IRAS, MSX) photometry.  
Table~\ref{primary1} shows the 2MASS $J$, $H$, and $K_S$ magnitudes \citep{twomassref}
and the derived extinction for each primary calibrator. 

The reddened stellar SED template for each primary calibrator
gives its flux density at all wavelengths on a scale consistent
with the network of standards constructed by \citet{cohenx} 
and absolutely validated by \citet{price}.
The fluxes in the IRAC wavebands were calculated by integrating the
normalized spectral template for each star over the IRAC spectral response.
The detailed derivation of the calibration factors is as follows.
The number of electrons per unit time collected from a source with spectrum 
$F_\nu$ using a telescope with area $A$ is
\begin{equation}
N_e = A \int{\frac{F_\nu}{h\nu} R d\nu},
\end{equation}
where $R$ is the system spectral response (in units of electrons per
photon, at frequency $\nu$, determined from pre-launch measurements). 
The calibration factor is the ratio of the flux density
at the nominal wavelength, $\lambda_0=c/\nu_0$, if the source had the
nominal $\nu F_\nu$=constant spectrum, to the observed electron 
production rate:
\begin{equation}
{\cal C'} = \frac{F_{\nu_0}^{*}K^*}{N_e},
\end{equation}
where the color correction
\begin{equation}
K \equiv \frac{\int{\left(F_\nu/F_{\nu_0}\right) \left(\nu/\nu_0\right)^{-1} R d\nu}}
              {\int{\left(\nu/\nu_0\right)^{-2} R d\nu}}.
\label{colcor}\end{equation}
For an imager, like IRAC, we calibrate in units of surface brightness,
and the flux measurements are made using aperture photometry in a finite aperture
that does not necessarily include all of the flux. 
We will continue to use $F_\nu$
as the true source spectrum, and define the calibration factor so that after
both aperture and color corrections to the observations are performed, the 
best measurement of the source flux is obtained. 
To convert from raw IRAC units of data numbers (DN) into surface brightness per
pixel, the calibration factor is:
\begin{equation}
{\cal C} = {\cal C'} \frac{G}{f_{ap} \Omega_{pix}} = 
\frac{F_{\nu_0}^{*}K^* G}{f_{ap} N_e \Omega_{pix}},
\end{equation}
where $G$ is the gain (electrons/DN) and $\Omega_{pix}$ is the pixel
solid angle (pixels are square with $1.22''$ sides), 
and $f_{ap}$ is the aperture correction factor (taken to
be unity; see \S\ref{aperture}); 
the units of the calibration factor ${\cal C}$ are
(MJy/sr)/(DN/s). The $*$ superscript indicates the quantity refers to
a calibration star, rather than a generic spectrum.
It is perhaps worth emphasizing that the only properties of the stellar
spectra that enter into the absolute calibration are the integrated
in-band fluxes for each IRAC channel, and the calibration factors
are directly proportional to the in-band flux.

\begin{figure}[th]
\epsscale{1}
\plotone{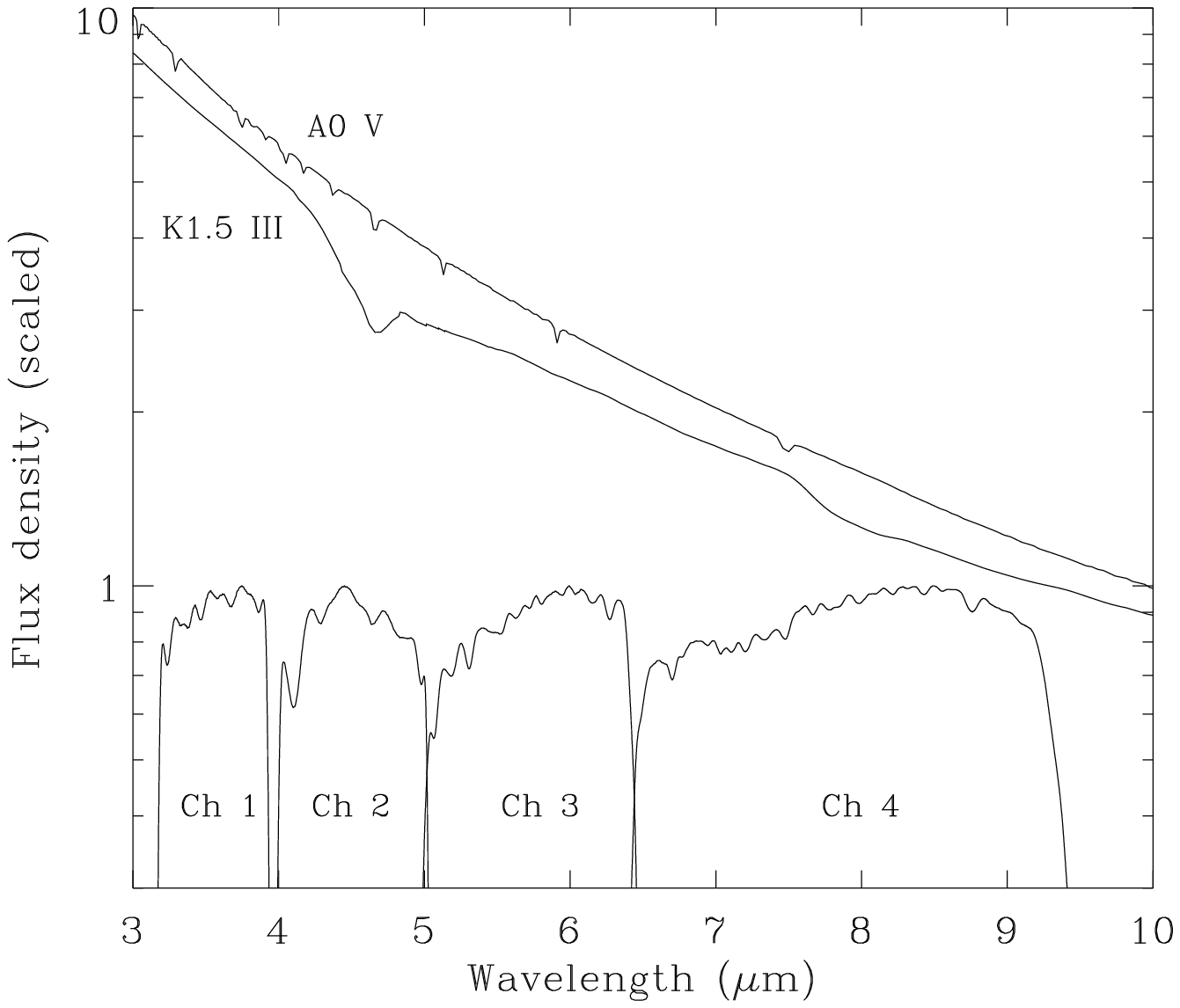}
\epsscale{1}
\figcaption[f7.eps]{Predicted spectra of two primary calibrators,
an A dwarf and a K giant, together with the relative spectra responses
of the four IRAC channels. Both the spectra (flux density units; $\propto$ Jy) 
and relative spectral response (in electrons per photon) were
scaled arbitrarily to unity to fit on this plot.
\label{calspec}}
\end{figure}

Table~\ref{primary2} ~shows the predicted fluxes of the IRAC primary calibrators.
The flux predictions, $F_{\nu_0}^{*}K^*$, are in IRAC `quoted' units,
such as would be obtained from science products from the SSC data analysis pipelines.
The monochromatic flux at the nominal wavelength can be determined
(for the calibrators or for any pipeline-processed IRAC data) by dividing
the `quoted' flux by the color correction (eq.~\ref{colcor}).
The nominal wavelengths for the IRAC bands were chosen to minimize the
color corrections. One can show (by making a Taylor series
expansion of the source spectrum) 
that the minimal dependence of the color correction on the slope of the
source spectrum is obtained for the nominal wavelength defined as follows:
\begin{equation}
\lambda_0 = 
\frac{\int{\lambda \nu^{-1} R d\nu}}
       {\int{\nu^{-1} R d\nu}}.\label{lambda0}
\end{equation}
The nominal wavelengths calculated in this way, using the
current spectral response curves \citep{fazio}, are
 3.550, 4.493, 5.731, and 7.872 $\mu$m in channels 1, 2, 3, and 4,
respectively. 
Figure~\ref{calspec} shows the model spectra for an A dwarf and 
K giant calibrator, together with the spectral responses.
We calculated the
calibrator color-corrections $K^*$ separately for each calibrator spectrum.
For reference, the color corrections are (in channels 1, 2, 3, and 4, respectively) 
for an A1V star:             $K^*_A=$1.019, 1.001, 1.026, and 1.042;
for an K1.5III star:         $K^*_K=$1.021, 1.070, 1.001, and 1.093;
for Rayleigh-Jeans spectra: $K_{RJ}=$1.011, 1.012, 1.016, and 1.034.

The IRAC spectral response calibration convention, 
such that no color corrections are needed for sources 
with $\nu F_\nu$=constant spectra,
is essentially the same convention used by
{\it IRAS} \citep{Beichman88}, 
{\it COBE}/DIRBE \citep{hauser98}, 
and {\it ISO} \citep{Siebenmorgen99},
except that the monochromatic flux densities 
are at the nominal wavelength (eq.~\ref{lambda0}) 
for each channel rather than a round-number wavelength.

\begin{deluxetable}{rlccccl}
\tabletypesize{\scriptsize}
\tablecaption{Predicted brightness of Primary absolute calibrators\label{primary2}}
\tablecolumns{7}
\tablehead{     &            &  \multicolumn{4}{c}{Predicted flux, $F_{\nu_0}^{*}K^*$ (mJy)\tablenotemark{a}}     & \\\cline{3-6}
\colhead{\#} & \colhead{Star} & \colhead{3.6 $\mu$m} & \colhead{4.5 $\mu$m}& \colhead{5.8 $\mu$m}& \colhead{8.0 $\mu$m} &\colhead{Campaign log}}
\startdata
1    &NPM1p64.0581&$\phn 63.03\pm \phn 1.49$    &$\phn 41.03\pm \phn 0.95$     &$\phn 26.17\pm  0.62$    &$\phn 14.40\pm  0.34$& yyyyyynnnny\\
2    &HD165459    &$647.38\pm 17.02$            &$421.19\pm 10.85$             &$268.71\pm  7.06$        &$148.09\pm  3.90$& yyyyyyyyyyy\\
3    &NPM1p60.0581&$\phn 38.20\pm \phn 0.89$    &$\phn 24.74\pm \phn 0.56$     &$\phn 15.75\pm  0.37$    &$\phn\phn  8.70\pm  0.20$& nnnnyyyyyyy\\
4    &1812095     &$\phn\phn  8.68\pm \phn 0.20$&$\phn\phn  5.66\pm\phn  0.12$ &$\phn\phn  3.62\pm 0.08$ &$\phn\phn  2.00\pm  0.04$& yyyyyyyyyyy\\
5    &KF08T3      &$\phn 11.78\pm \phn 0.42$    &$\phn\phn  7.25\pm \phn 0.45$ &$\phn\phn  4.64\pm 0.13$ &$\phn\phn  2.69\pm  0.08$& yyyyyyyyyyy\\
6    &KF09T1      &$169.94\pm \phn 5.86$        &$104.69\pm \phn 6.33$         &$\phn 67.02\pm  1.74$    &$\phn 38.76\pm  1.04$& ynnnyyyyyyy\\
7    &KF06T1      &$\phn 13.93\pm \phn 0.46$    &$\phn\phn  7.80\pm \phn 0.50$ &$\phn\phn  5.34\pm  0.13$&$\phn\phn  3.09\pm  0.08$& yyyyyynnnny\\
8    &KF06T2      &$\phn 10.53\pm \phn 0.35$    &$\phn\phn  5.99\pm \phn 0.38$ &$\phn\phn  4.05\pm  0.10$&$\phn\phn  2.34\pm  0.06$& nnnnyyyyyyy\\
9    &NPM1p66.0578&$140.94\pm \phn 6.11$        &$\phn 82.39\pm \phn 7.34$     &$\phn 54.54\pm  1.60$    &$\phn 29.72\pm  1.03$& yyyyyynnnny\\
10   &NPM1p67.0536&$843.49\pm 35.09$            &$482.35\pm 39.32$             &$319.84\pm  9.81$        &$185.25\pm  7.39$& yyyyyyyyyyy\\
11   &NPM1p68.0422&$580.54\pm 24.89$            &$335.64\pm 27.73$             &$223.18\pm  7.01$        &$128.96\pm  5.22$& yyyyyyyyyyy\\
\enddata
\tablenotetext{a}{Predicted flux density in `quoted' IRAC convention. The predicted flux
density at the nominal wavelength is $F_{\nu_0}^{*}$, and the color correction for
each star's predicted spectrum spectrum is $K^*$.}
\end{deluxetable}

\subsection{Comparison between different standard stars}

\begin{figure}[th]
\epsscale{1}
\plotone{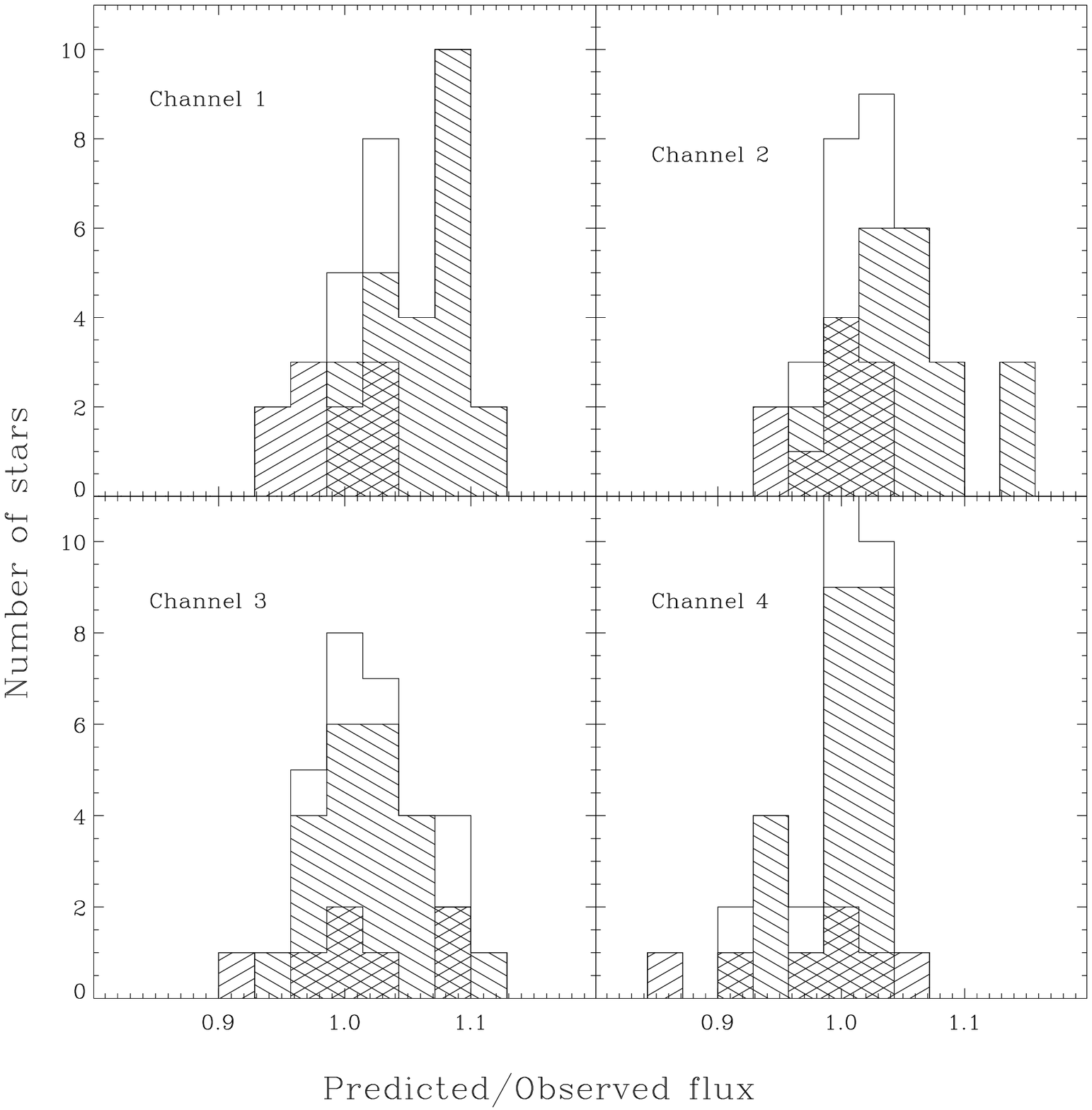}
\epsscale{1}
\figcaption[f8.eps]{Distributions of predicted/observed
fluxes for the 34 calibrator candidates in each IRAC channel.
Each box represents one star. The histogram of the total sample is
the upper envelope. The histogram for A stars only has diagonal
hatching one way (/), and the histogram for K stars only has diagonal
hatching another way ($\backslash$). The two distributions overlap near unity, but
the displacement is evident, with the K stars having higher predicted
than observed flux.
\label{calstar_ioc}}
\end{figure}

To search for systematic trends in the comparison between 
observed and predicted fluxes, we used two
samples: the primary standard stars selected for the nominal mission 
(Table~\ref{primary1}; 11 stars) and the
candidate calibrators observed during the in-orbit checkout
(Table~\ref{calstar_ioc_table}; 34 stars). The primary standards
defined the IRAC calibration, so we consider the candidate
calibrators as a relatively independent comparison sample
to search for trends. Figure~\ref{calstar_ioc} shows the 
histograms of the predicted/observed fluxes for the candidate
calibrators, for each IRAC channel. The `observed' fluxes 
have been scaled using the calibration factors averaged
over the primary standards in the method described below
in \S\ref{newcalfac}. The distributions are centered near
unity, demonstrating good agreement between that the candidate 
calibrator and primary standards. 

We found one systematic trend in the predicted/observed fluxes:
a displacement between the calibration derived from A dwarfs and K 
giants, in IRAC channels 1 and 2. Figure~\ref{calstar_ioc} shows
the histogram of the calibration factors separately for these two types
of primary calibrators. The difference in average calibration inferred
from the two types of calibrators is twice as large than the
widths (3.4\%) of the distributions for each calibrator type. 
This leads to an unnecessary systematic
error in the calibration, with a spectral dependence that
changes channels 1 and 2 with respect to the other IRAC channels. 
To bolster the statistics, we include the primary standards together
with the calibrators observed only in IOC; the total sample
comprises 13 A stars and 29  K stars.
The difference between the A and K star distributions of predicted/observed
fluxes is 
$7.3\pm 2.3$\% in channel 1,
$6.5\pm 2.4$\% in channel 2,
$3.6\pm 2.5$\% in channel 3, and
$2.1\pm 2.8$\% in channel 4,
with the predicted/observed higher for K stars than for A stars in all
four channels.
Our original calibration strategy was to observe both A and K calibrators,
then take the average to guard against model uncertainties. 
After seeing such a clear systematic difference, with a color dependence,
between the two types of calibrators, and furthermore seeing that there
is no temporal drift in the calibration (so that we can use all calibration
star data all the way back to the in-orbit checkout), we 
decided to adopt an A-star-only absolute calibration convention. 
At present, we do not know why the K giants are systematically offset
from the A stars, nor do we conclusively know which of the two
types of calibration are correct.
With the relatively small number of stars
observed and analyzed here, and only two types of calibrators
analyzed, it is not
possible to conclusively say whether the template spectra
are inaccurate representations of the stars, 
or the particular sample of stars is anomalous.

The distribution of predicted/observed flux
(Figure~\ref{calstar_ioc}) is non-gaussian, because
it is a combination of the measurement uncertainties (which do have
a gaussian distribution) and the systematic uncertainties (which
do not). In all four IRAC channels (but particularly channel 1), 
two peaks in the distribution are evident due to the separate
distributions of A and K stars. 
Further, there are too few calibrators for the central limit theorem to 
lead to a clean gaussian distribution for each type. 
The dispersions of the distributions are $\sim 2.5-3.5$\%, which is
comparable to the uncertainty in the predicted fluxes based
only on pre-launch ground and space measurements with other telescopes.

\subsection{Best calibration factors and their uncertainties\label{newcalfac}}

The recommended calibration factors for IRAC were calculated
as a weighted average of calibration factors for the four, A-type
primary calibrators.  The results are given in 
Table~\ref{newcaltab}\footnote{The calibration factors in Table~\ref{newcaltab}
have been applied to IRAC data with pipeline version S11.}.
Weights were inversely proportional to the uncertainties in the absolute
calibration of each star \citep{cohenxiii}.

The total uncertainty in the calibration factor is {\it not}
the statistical uncertainty in the weighted 
mean, since $\sigma_{abs}=1.5$\% of the uncertainty in each of the predicted 
fluxes arises from the absolute calibration of Vega and Sirius, which 
scales all of the predictions. 
To assess the uncertainty in the calibration, we separate
three items in Table~\ref{newcaltab}.
The absolute calibration uncertainty for each star is $\sigma_{ground}$.
The dispersion in calibration factors derived from an individual star
is $\sigma_{rms}$, attributed largely to measurement errors.
Finally, the campaign-averaged calibration factor for each
star has each calibrator has some dispersion from campaign to
campaign of $\sigma_{repeat}$, which limits long-term drifts
(which are apparently not present for IRAC).
The combined uncertainty for $n$ calibrators determined from
\begin{equation}
\sigma_{\cal C}=\sqrt{\sigma_{abs}^2+(\sigma_{ground}^2-\sigma_{abs}^2)/n
+\sigma_{rms}^2/n}.
\end{equation}
We used this equation, with $n=4$, to derive the uncertainty 
(2.0\% in all four IRAC bands) in the
absolute calibration factors listed in Table~\ref{newcaltab}.

\begin{table}
\caption[h]{Absolute calibration factors for IRAC$^a$}\label{newcaltab} 
\begin{flushleft} 
\begin{tabular}{lllll} 
\hline
Channel & Calibration factor$^a$ & $\sigma_{ground}$ & $\sigma_{rms}$
 & $\sigma_{repeat}$\\\hline\hline
1     & $0.1088\pm 0.0022$ & 2.3\% & 2.0\% & 0.9\% \\
2     & $0.1388\pm 0.0027$ & 2.3\% & 1.9\% & 0.9\% \\
3     & $0.5952\pm 0.0121$ & 2.3\% & 2.1\% & 0.9\% \\
4     & $0.2021\pm 0.0041$ & 2.3\% & 2.1\% & 0.8\% \\
\hline
\end{tabular} 
\end{flushleft} 
$^a$ units: (MJy/sr)/(DN/s)
\end{table}  

Including the larger sample of calibrators observed only during IOC, 
the calibration differs from the values in Table~\ref{newcaltab}
by -0.9, -1.1, -1.7, and -2.3 \%, respectively, in channels 1, 2, 3, and 4.
This difference is marginally significant and we incorporate it into
the overall calibration uncertainty.
The IRAC calibration will be updated in the future to include these
results and a wider sample of spectral types from the Legacy science
programs.

\subsection{IRAC magnitude system}

We define the IRAC magnitude system such that an observer measures
the flux density, $F_\nu^{quot}$, of a source from the calibrated images
out of the IRAC pipeline, performs image-based corrections 
(array-location-dependent
photometric correction, pixel-phase correction, aperture correction),
and then uses the zero-magnitude flux densities, $F_{zero}^{[i]}$,
to calculate the magnitude $[i]=2.5\log_{10}(F_{zero}^{[i]}/F_\nu^{quot})$
where $i=3.6, 4.5, 5.8, 8$ are the four IRAC channels. In this 
system, there is no need to know the spectral shape of the source,
as the magnitude is a measure of the in-band flux relative to that of Vega.
The zero-magnitude fluxes were determined by integrating the
Kurucz model spectrum of $\alpha$ Lyr over the passbands using
the equations above; specifically,
\begin{equation}
F_{zero}^{[i]} = \frac{\int{F_\nu \left(\nu/\nu_0\right)^{-1} R^{[i]} d\nu}}
              {\int{\left(\nu/\nu_0\right)^{-2} R^{[i]} d\nu}}.
\end{equation}
The resulting zero-magnitude flux densities are
$280.9 \pm 4.1$,	$179.7 \pm 2.6$,	$115.0 \pm 1.7$, and $64.13 \pm 0.94$ Jy 
in the [3.6], [4.5], [5.8], and [8] $\mu$m channels, respectively.
Due to the choice of A dwarfs as the absolute calibrators, this magnitude
convention should yield results on the same scale as used in 
optical and near-infrared astronomy.

\subsection{Aperture corrections and extended emission\label{aperture}}

All the discussion so far has been based on aperture photometry
with a specified beam size and sky annulus (\S\ref{photometrysection}), 
and the
calibration factors in Table~\ref{newcaltab} are defined for these parameters.
The parameters were optimized for photometry of isolated, bright
point sources and will not in general be suitable for other
applications.  In particular, some of the source flux lands in
the sky annulus and is subtracted, while additional flux lands
outside the sky annulus and is ignored.
Measurements using different beam sizes or sky annuli will need
to account for these effects.  Therefore,
\begin{enumerate}
\item Point sources extracted with the same aperture as we used for calibration 
will get the correct flux (no further aperture correction needed);
\item Point sources extracted with other on-source radii or sky annuli can
use aperture correction tables, as long as they are normalized such that
there is no correction for apertures identical to those used in our
absolute calibration;
\item Point sources extracted using point spread function fitting should verify
the normalization of the point spread function would give unity flux if
the on-source and sky calibration apertures are applied (i.e. not unity
flux for the integral over the entire point spread function); and
\item extended emission surface brightness will be incorrectly calibrated
and will require scaling by the "infinite" aperture correction.
\end{enumerate}

The reason we have not attempted an aperture correction is that the measurements
are still under way and the empirically-derived values are not well understood.
Part of the effect is simple diffraction and could be estimated using, for
example, an Airy function determined by the primary mirror size and the
filter central wavelength; this effect explains channels 1 and 2. However, in
channels 3 and 4 substantially more of the source flux is scattered out of
the calibration aperture than can be explained by diffraction theory. This
light is thought to be scattered within the detector material. Its distribution,
dependence on location of the source in the array, and the ultimate fate of the 
lost source flux are under investigation. The empirical determination of the aperture
corrections is being described by \citet{marengo}; at present we recommend extended
emission (including the basic calibrated data and the image mosaics from
the pipeline) be multiplied by the effective aperture correction factors of
0.944, 0.937, 0.772, and 0.737 for channels 1, 2, 3, and 4, respectively.

Three further practical matters have already been mentioned but
are worth summarizing here as well. 
\begin{enumerate}
\item The IRAC science pipeline generates
images in units of surface brightness, but
because of distortion, the pixels do not subtend constant solid
angle. 
\item The IRAC spectral response varies over the field of view, and
therefore the color corrections are field dependent.  
\item The electron rate in the 3.6 $\mu$m channel depends slightly on
pixel phase.  The present calibration factors is correct on average,
as is appropriate for sources observed multiple times at multiple
dither positions.  For the most precise photometry, however, pixel
phase should be taken into account.  
\end{enumerate}
Corrections (1) and (2) are available from the Spitzer Science Center
and are described by \citet{horafov}; they were taken
into account in the present paper by using the array-location-dependent
photometric corrections in Table~\ref{delta_flat}.
Correction (3) was applied using the simple equations in \S\ref{photometrysection}.
Observers should apply these corrections in a manner
consistent with that applied in this paper,
in order for the absolute calibration to apply.

\section{Serendipitous results}

\begin{figure}[th]
\epsscale{1}
\plotone{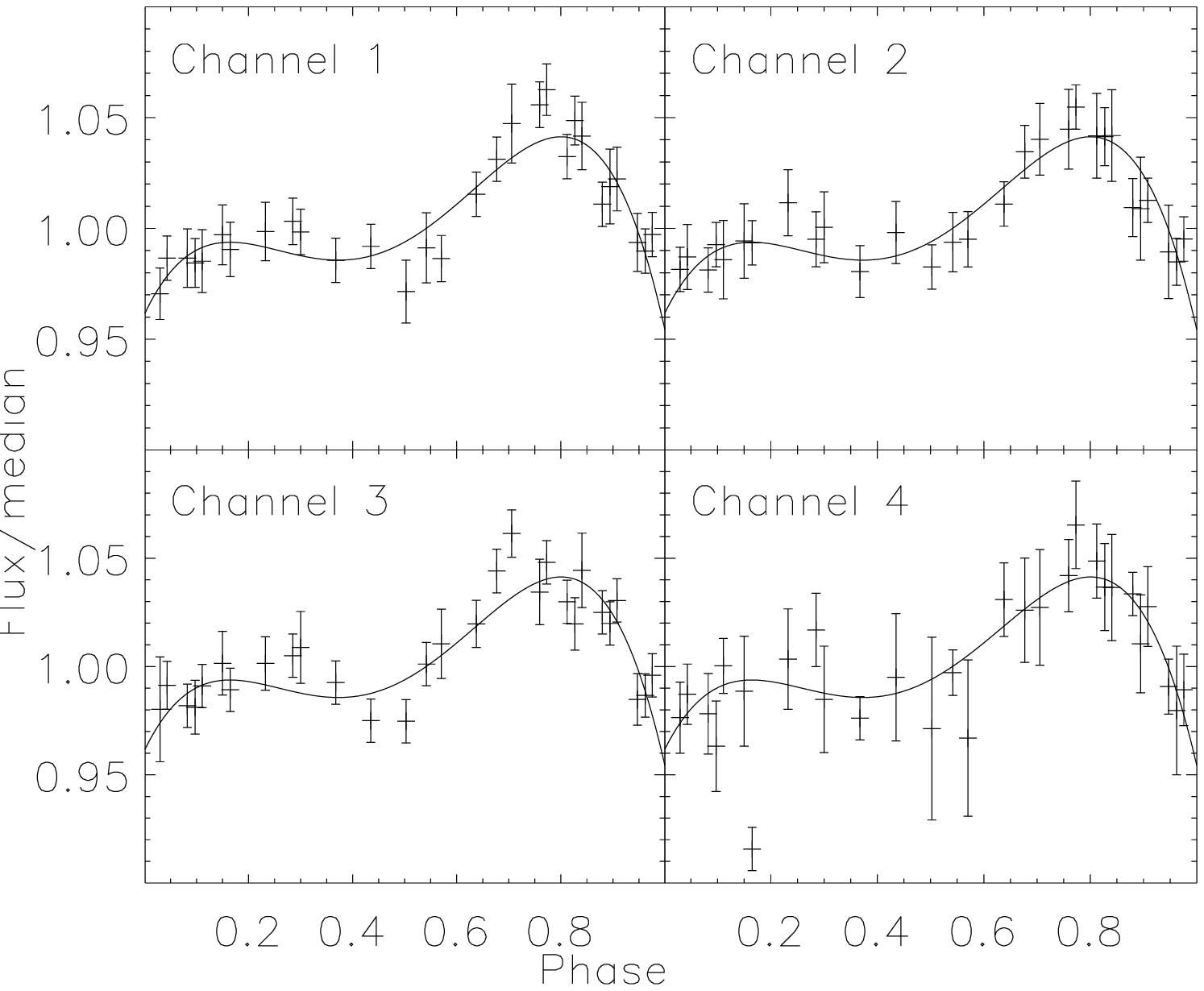}
\epsscale{1}
\figcaption[f9.eps]{Light curve of the secondary calibrator SA115\_554 in
the four IRAC channels. The solid curve (same exact curve in each of the
four panels) is a fourth-order polynomial fit to all of the wavelengths,
indicating the light curve is color-invariant.
\label{eclip}}
\end{figure}

The secondary calibrator SA115\_554 has a periodic light curve. It was observed
every 12 hr in two campaigns, spanning a total of 26 days. The light-curve phases
well to a period of 7.4 days, with an amplitude of 5\%. Figure~\ref{eclip} shows
the light curve, phased to 0 at modified Julian date 52996.5. The light curve 
is identical at all 4 IRAC wavelengths, indicating that the source of variability 
is not a mid-infrared-specific phenomenon but rather something with colors 
comparable to the star, which is classified as a K1.5III. This star was found
to be variable as part of the ASAS by \citet{pojmanski}, who found
I-band variability with a period of 7.451 days, identical to the 
IRAC period. The amplitude of the I-band variation was 0.18 mag, and
its light-curve is distinctly different from that measured by IRAC. 
If the variability were due to an orbiting companion, then
the companion would probably be late-type and small, perhaps an M dwarf.
But the observed variability likely arises from pulsations of the star itself.
The optical light curve is more characteristic of a Cepheid variable,
and the spectral type obtained by \citet{drilling} was G7, consistent
with a Cepheid, though the spectral type derived from our own work
was a K1.5III.


The primary calibrator HD 165459 was found to have a significant ($\sim 40$\%)
excess at 24 $\mu$m, in the MIPS calibration program.
Presumably, this excess is due to a disk around the star such
as in other `Vega'-type A stars. Is there any evidence for this
disk at IRAC wavelengths? Calibrating the measured fluxes
with the new calibration factors, we find the star's flux relative
to the photospheric model is 
$0.982\pm 0.020$, $0.987\pm 0.024$, 
$0.976\pm 0.022$, and $0.993\pm 0.022$, 
in channels 1, 2, 3, and 4, respectively. 
Since these are absolute measurements,
we have used $\sigma_{rms}$ from Table~\ref{newcaltab} as the 
calibration uncertainty (and combined with the uncertainty in this
individual star's measurements).
In fact, we can remove the absolute uncertainty in
the size of the star by normalizing
all fluxes at 3.6 $\mu$m. Then the ratio of observed flux to photospheric, 
assuming there is no excess
at 3.6 $\mu$m, is $1.006\pm 0.009$, $0.994\pm 0.009$, and $1.011\pm 0.008$
in channels 2, 3, and 4, respectively.
It appears that there is no mid-infrared excess for
this star in the 3.6--8 $\mu$m range, at a 95\% confidence upper
limit of 4\% (with no assumptions),
or at 4.5, 5.8, or 8 $\mu$m, at a 95\% confidence limit of 
1.2, 1.8, or 1.6\% (assuming no excess at 3.6 $\mu$m).

\section{Conclusions}

The Infrared Array Camera on the {\it Spitzer Space Telescope} has a stable
gain on all measured timescales, making it possible to measure variability
at the 2\% level for carefully-reduced data. The absolute calibration using
stellar photospheric models scaled to ground-based photometry at optical 
through near-infrared wavelengths is accurate to 1.8\%, 1.9\%, 2.0\%,
and 2.1\% in channels 1 (3.6 $\mu$m), 2 (4.5 $\mu$m), 
3 (5.8 $\mu$m), and 4 (8 $\mu$m), respectively. To measure fluxes at
this high level of accuracy requires several photometric corrections:
array position dependence (due to changing spectral response and
pixel solid angle over the camera of view), pixel phase dependence (due
to nonuniform quantum efficiency over a pixel), color correction 
(due to the different system response integrated over the passband for
sources of different color), and aperture correction (due to the fractions
of light included within the measurement aperture and lost in the
background aperture).

\acknowledgements 

This work is based 
on observations made with the Spitzer Space Telescope, which is operated
by the Jet Propulsion Laboratory, California Institute of Technology under NASA
contract 1407. Support for this work was provided by NASA through an award
issued by JPL/Caltech. 
This publication makes use of data products from the Two
Micron All Sky Survey, which is a joint project of the University of
Massachusetts and the Infrared Processing and Analysis Center/California
Institute of Technology, funded by the National Aeronautics and Space
Administration and the National Science Foundation. 
We thank Don Hoard and Stephanie Wachter for helping us understand the
nature of the variable secondary calibrator and pointing out that it might be a
Cepheid variable. 
MC thanks SAO for support under prime contract
SV9-69008 with UC Berkeley.


\clearpage

\begin{thebibliography}{} 
\bibitem[Beichman et al.(1988)]{Beichman88} Beichman, C. et al. 1988. 
{\it Infrared Astronomical
Satellite (IRAS): Catalogs and Atlases, Volume 1. Explanatory
Supplement} (NASA RP-1190), \S VI.C.6

\bibitem[Cohen(2003)]{mcisocal}
Cohen, M. 2003, 
Proc. ``The Calibration Legacy of the ISO Mission", ed. L. Metcalfe,
        VILSPA, Spain (ESA SP-481), pg.135

\bibitem[Cohen et al.(2003)]{cohenxiii} Cohen M., Megeath, S. T., Hammersley, P. L.,
Mart\'in-Luis, F., \& Stauffer, J. 2003. 
AJ, 125, 2645

\bibitem[Cohen, Wheaton, \& Megeath(2003)]{cohenxiv} Cohen M., Wheaton, Wm. A.,
\& Megeath, S. T. 2003. 
AJ, 126, 1090

\bibitem[Cohen et al.(1999)]{cohenx} Cohen M., Walker, R. G., Carter, B., Hammersley, P.,
Kidger, M., \& Noguchi, K. 1999
AJ, 117, 1864


\bibitem[Cohen et al.(1996)]{cohenvii}
Cohen, M., Witteborn, F. C., Carbon, D. F., Davies, J. K.,
Wooden, D. H., Bregman, J. D. 1996, AJ, 112, 2274

\bibitem[Drilling \& Landolt(1979)]{drilling} Drilling, J. S., \& 
Landolt, A. U. 1979. AJ, 84, 783

\bibitem[Fazio et al.(2004)]{fazio}
Fazio, G. G. et al. 2004, ApJS, 154, 10

\bibitem[Hauser et al.(1998)]{hauser98} Hauser, M. G. {\it et al.} 1998. 
{COBE Diffuse Infrared Background Experiment (DIRBE) Explanatory Supplement}, 
\S 5.5, see http://borneo.gsfc.nasa.gov/cobe1/DIRBE/DOC/ 

\bibitem[Hora et al.(2005)]{horafov} Hora, J. et al. 2005, in preparation

\bibitem[Klemola et al.(1987)]{klemola} Klemola, A. R., Jones, B. F., \&
Hanson, R. B. 1987, AJ, 94, 501 

\bibitem[K\"ummel \& Wagner(2000)]{kummel} K\"ummel, M. W. \& Wagner, S. J.
2000, A\&A, 353, 867

\bibitem[Kurucz(1993)]{kurucz}
Kurucz, R.L. 1993, CD-ROM series; ``ATLAS9 Stellar Atmosphere Programs and 2 km/s grid",
CD-ROM No.13); ``SYNTHE Spectrum Synthesis Programs and Line Data" (Kurucz CD-ROM No.118)

\bibitem[Marengo et al.(2005)]{marengo} Marengo, M. et al. 2005, in preparation

\bibitem[Megeath et al.(2003)]{stmisocal} 
Megeath, S.T., Cohen, M., Stauffer, J., Hora, J., Fazio G., Berlind, P.,
\& Calkins, M. 2003, 
Proc. ``The Calibration Legacy of the ISO Mission", ed. L. Metcalfe, VILSPA, 
Spain (ESA SP-481), pg.165

\bibitem[Pojma\'nski(2000)]{pojmanski} Pojma\'nski, G. 2000.
Acta Astronomica, 50, 177.

\bibitem[Price et al.(2004)]{price}
Price, S.D., Paxson, C., Engelke, C., Murdock, T.L. 2004, AJ, 128, 889

\bibitem[Siebenmorgen et al.(1999)]{Siebenmorgen99} Siebenmorgen, R., 
Blommaert, J., Sauvage, M., \& Starck, J.-L. 1999.
{\it ISO Handbook Volume III (CAM), Version 1.0}, \S 4.1.1, see
http://isowww.estec.esa.nl/manuals/HANDBOOK/III/cam\_hb/ 

\bibitem[Skrutskie(1999)]{twomassref} Skrutskie, M. F. 1999, in {\it Astrophysics with Infrared
Surveys: A Prelude to SIRTF}, eds. M. D. Bicay, C. A. Beichman, R. M. Cutri, \&
B. F. Madore (San Francisco: ASP), p. 185.

\bibitem[Werner et al.(2004)]{werner}
Werner, M. W. et al. 2004, ApJS, 154, 1

\end{thebibliography}
\end{document}